\documentclass[11pt,preprint]{aastex}
\pdfoutput=1
\usepackage{graphicx}

\let\oldfootsep=\footnotesep
 \def\VEV#1{\left\langle #1\right\rangle}
\newcommand\ltsima{$\; \buildrel <\over\sim \;$}
\newcommand\simlt{\lower.5ex\hbox{\ltsima}}
\newcommand\gtsima{$\; \buildrel >\over\sim \;$}
\newcommand\simgt{\lower.5ex\hbox{\gtsima}}

\newcommand\msun {M_\odot}

\newcommand\mearth {{M_\oplus}}

\newcommand\pac{Paczy{\'n}ski }
\newcommand\ie{{i.e.}}

\newcommand{\mathbold}[1]{\mbox{\boldmath $\bf#1$}}
\newcommand\piEbold{{\mathbold \pi_E}}
\newcommand\rep {\tilde{r}_E}
\newcommand\repbold {{\mathbold \rep}}
 
\shorttitle{}
\shortauthors{Bennett et al}

\begin{document}


\title{Masses and Orbital Constraints for the OGLE-2006-BLG-109Lb,c \\
         Jupiter/Saturn Analog Planetary System}


\author{D.P.~Bennett\altaffilmark{1},\\
S.H.~Rhie\altaffilmark{1},
S.~Nikolaev\altaffilmark{2},\\
B.S.~Gaudi\altaffilmark{3},
A.~Udalski\altaffilmark{4},
A.~Gould\altaffilmark{3},
G.W.~Christie\altaffilmark{5},
D.~Maoz\altaffilmark{6},
S.~Dong\altaffilmark{3}, \\
J.~McCormick\altaffilmark{7},
M.K.~Szyma{\' n}ski\altaffilmark{4}, 
P.J.~Tristram\altaffilmark{8}, 
B.~Macintosh\altaffilmark{2},
K.H.~Cook\altaffilmark{2}, \\ \ \\
M.~Kubiak\altaffilmark{4}, 
G.~Pietrzy{\' n}ski\altaffilmark{4,9}, 
I.~Soszy{\' n}ski\altaffilmark{4},
O.~Szewczyk\altaffilmark{9},
K.~Ulaczyk\altaffilmark{4}, 
{\L}.~Wyrzykowski\altaffilmark{4,10}\\ 
(The OGLE Collaboration)\\
D.L.~DePoy\altaffilmark{11}, 
C.~Han\altaffilmark{12}, 
S.~Kaspi\altaffilmark{6}, 
C.-U.~Lee\altaffilmark{13}, 
F.~Mallia\altaffilmark{14},\\
T.~Natusch\altaffilmark{15},
B.-G.~Park\altaffilmark{13},
R.W.~Pogge\altaffilmark{3}, 
D.~Polishook\altaffilmark{6}, \\
(The $\mu$FUN Collaboration)\\
F.~Abe\altaffilmark{16}, 
I.A.~Bond\altaffilmark{17}, 
C.S.~Botzler\altaffilmark{18}, 
A.~Fukui\altaffilmark{16}, 
J.B.~Hearnshaw\altaffilmark{19},
Y.~Itow\altaffilmark{16},  
K.~Kamiya\altaffilmark{16}, 
A.V.~Korpela\altaffilmark{20},
P.M.~Kilmartin\altaffilmark{8}, 
W.~Lin\altaffilmark{17},
J.~Ling\altaffilmark{17},
K.~Masuda\altaffilmark{16},  
Y.~Matsubara\altaffilmark{16}, 
M.~Motomura\altaffilmark{16}, 
Y.~Muraki\altaffilmark{21}, 
S.~Nakamura\altaffilmark{16},
T.~Okumura\altaffilmark{16}, 
K.~Ohnishi\altaffilmark{22}, 
Y.C.~Perrott\altaffilmark{18}, 
N.J.~Rattenbury\altaffilmark{18}, 
T.~Sako\altaffilmark{16}, 
To.~Saito\altaffilmark{23},
S.~Sato\altaffilmark{24}, 
L.~Skuljan\altaffilmark{18}, 
D.J.~Sullivan\altaffilmark{20}, 
T.~Sumi\altaffilmark{16}, 
W.L.~Sweatman\altaffilmark{17},
P.C.M.~Yock\altaffilmark{18}, \\
(The MOA Collaboration)\\
M.~Albrow\altaffilmark{19}, 
A.~Allan\altaffilmark{25}, 
J.-P.~Beaulieu\altaffilmark{26}, 
D.M.~Bramich\altaffilmark{27},
M.J.~Burgdorf\altaffilmark{28,29}, 
C.~Coutures\altaffilmark{26},
M.~Dominik\altaffilmark{30}, 
S.~Dieters\altaffilmark{26}, 
P.~Fouqu\'e\altaffilmark{31}, 
J.~Greenhill\altaffilmark{32},
K.~Horne\altaffilmark{30},
C.~Snodgrass\altaffilmark{33},
I.~Steele\altaffilmark{34}, 
Y.~Tsapras\altaffilmark{35},\\
(From the PLANET and RoboNet Collaborations)\\
B.~Chaboyer\altaffilmark{36}, 
A.~Crocker\altaffilmark{37}, and
S.~Frank\altaffilmark{3}
              } 




\altaffiltext{1}{Department of Physics,
    University of Notre Dame, Notre Dame, IN 46556, USA; 
    Email: {\tt bennett@nd.edu}}
    
\altaffiltext{2}{IGPP, Lawrence Livermore National Laboratory, 7000 East Ave., Livermore, CA 94550, USA}
\altaffiltext{3}{Department of Astronomy, Ohio State University, 140 West 18th Avenue, Columbus, OH 43210, USA}
\altaffiltext{4}{Warsaw University Observatory, Al.~Ujazdowskie~4, 00-478~Warszawa, Poland}
\altaffiltext{5}{Auckland Observatory, P.O. Box 24-180, Auckland, New Zealand}
\altaffiltext{6}{School of Physics and Astronomy, Faculty of Exact Sciences, Tel-Aviv University, Tel Aviv 69978, Israel}
\altaffiltext{7}{Farm Cove Observatory, 2/24 Rapallo Place, Pakuranga, Auckland 1706, New Zealand}
\altaffiltext{8}{Mt. John Observatory, P.O. Box 56, Lake Tekapo 8770, New Zealand}
\altaffiltext{9}{Universidad de Concepci{\'o}n, Departamento de Fisica, Casilla 160--C, Concepci{\'o}n, Chile}
\altaffiltext{10}{Institute of Astronomy, University of Cambridge, Madingley Road, Cambridge CB3 0HA, UK}
\altaffiltext{99}{Department of Physics \& Astronomy, Texas A\&M University, 
College Station, TX 77843}
\altaffiltext{12}{Department of Physics, Chungbuk Nat.~University, 410 Seongbong-Rho, Hungduk-Gu, Chongju 371-763, Korea}
\altaffiltext{13}{Korea Astronomy and Space Science Institute, 61-1 Hwaam-Dong, Yuseong-Gu, Daejeon 305-348, Korea}
\altaffiltext{14}{Campo Catino Astronomical Observatory, P.O. Box Guarcino, Frosinone 03016, Italy}
\altaffiltext{15}{AUT University, School of Computing and Mathematical Sciences, Auckland, New Zealand}
\altaffiltext{16}{Solar-Terrestrial Environment Laboratory, Nagoya University, Nagoya, 464-8601, Japan}
\altaffiltext{17}{Institute for Information and Mathematical Sciences, Massey University, Auckland 1330, New Zealand}
\altaffiltext{18}{Department of Physics, University of Auckland, Private Bag 92-019, Auckland 1001, New Zealand}
\altaffiltext{19}{University of Canterbury, Department of Physics and Astronomy, Christchurch 8020, New Zealand}
\altaffiltext{20}{School of Chemical and Physical Sciences, Victoria University, Wellington, New Zealand}
\altaffiltext{21}{Department of Physics, Konan University, Nishiokamoto 8-9-1, Kobe 658-8501, Japan}
\altaffiltext{22}{Nagano National College of Technology, Nagano 381-8550, Japan}
\altaffiltext{23}{Tokyo Metropolitan College of Aeronautics, Tokyo 116-8523, Japan}
\altaffiltext{24}{Department of Physics and Astrophysics, Faculty of Science, Nagoya University, Nagoya 464-8602, Japan}
\altaffiltext{25}{School of Physics, University of Exeter, Stocker Road, Exeter, EX4 4QL, UK}
\altaffiltext{26}{Institut d'Astrophysique de Paris, CNRS, 98bis Boulevard Arago, 75014 Paris, France}
\altaffiltext{27}
{European Southern Observatory, Karl-Schwarzschild-Stra\ss{}e 2, 85748 Garching bei M\"unchen, Germany}
\altaffiltext{28} {SOFIA Science Center, NASA Ames Research Center, Mail Stop N211-3, Moffett Field CA 94035, USA}
\altaffiltext{29}{Deutsches SOFIA Institut, Universitaet Stuttgart, Pfaffenwaldring 31, 70569 Stuttgart, Germany}
\altaffiltext{30}{SUPA, University of St Andrews, School of Physics \& Astronomy,North Haugh, St Andrews, KY16 9SS, UK}
\altaffiltext{31}{LATT, CNRS, UniversitŽ de Toulouse, 14 avenue Edouard Belin, 31400 Toulouse, France}
\altaffiltext{32}{University of Tasmania, School of Mathematics and Physics, Private Bag 37, Hobart, TAS 7001, Australia}
\altaffiltext{33}
{European Southern Observatory, Casilla 19001, Vitacura 19, Santiago, Chile}
\altaffiltext{34}{Astrophysics Research Institute, Liverpool John Moores University, Egerton Wharf, Birkenhead CH41 1LD, UK}
\altaffiltext{35}{Las Cumbres Observatory, 6740B Cortona Dr, suite 102, Goleta, CA 93117, USA}
\altaffiltext{36}{Department of Physics and Astronomy, Dartmouth College, 6127 Wilder Laboratory, Hanover, NH 03755, USA}
\altaffiltext{37}{Department of Astronomy, University of Massachusetts, Amherst, MA 01002, USA}

\clearpage

\begin{abstract}
We present a new analysis of the Jupiter+Saturn analog system,
OGLE-2006-BLG-109Lb,c, which was the first double planet 
system discovered with the gravitational microlensing method.
This is the only multi-planet system discovered by any method
with measured masses for the star and both planets.
In addition to the signatures of two planets, this event also
exhibits a microlensing parallax signature 
and finite source effects that provide a direct
measure of the masses of the star and planets, and the expected
brightness of the host star is confirmed by Keck AO imaging,
yielding masses of $M_\ast = 0.51{+0.05\atop -0.04}\,\msun$,
$M_b = 231\pm 19\,\mearth$, and
$M_c = 86\pm 7\,\mearth$. 
The Saturn-analog planet in this system had a planetary light curve
deviation that lasted for 11 days, and as a result, the effects of the orbital
motion are visible in the microlensing light curve. We find that four
of the six orbital parameters are tightly constrained and that a fifth
parameter, the orbital acceleration, is weakly constrained. No orbital
information is available for the Jupiter-analog planet, but its presence
helps to constrain the orbital motion of the Saturn-analog planet.
Assuming co-planar orbits, we find an orbital eccentricity of
$\epsilon = 0.15 {+0.17\atop -0.10}$ and an orbital inclination of
$i = 64^\circ {+4^\circ \atop -7^\circ}$. The 95\% confidence
level lower limit on the inclination of $i > 49^\circ$ implies that 
this planetary system can be detected and studied via radial 
velocity measurements using a telescope of $\simgt 30\,$m
aperture.
\end{abstract}
\clearpage


\section{Introduction}
\label{sec-intro}
The discovery of the Jupiter/Saturn analog planetary system, OGLE-2006-BLG-109Lb,c,
by gravitational microlensing \citep{gaudi-ogle109} suggests that solar systems
like our own may be common, at least among systems that contain gas giant
planets. This was generally assumed to be the case prior
to the detection of the first extrasolar planets orbiting main sequence
stars \citep{51peg,70vir,47uma}, but
these first discoveries challenged this conventional wisdom, with the discovery
of hot-Jupiters and gas giants in highly eccentric orbits. 
However, the discovery of
true solar system analogs with the radial velocity method is difficult, requiring 
radial velocity precision $< 3\,{\rm m/s}$ spanning a decade or more
\citep{jup-twin}, so it may be that the apparent paucity of systems like our
own is a selection effect.

The gravitational microlensing method has very different selection effects
from the radial velocity method. It is most sensitive to planets at separations
similar to the Einstein radius, which in most cases is just beyond the
``snow line" at $\sim 2.7 M/\msun$
\citep{ida_lin,lecar_snowline,kennedy-searth,kennedy_snowline}
where gas giant planets are expected to form, according to the 
core accretion theory. Microlensing does find that super-Earth and Neptune-mass
planets \citep{ogle390,ogle169,bennett-moa192} are more common at
these separations than Jupiters. Of the handful of microlensing events that reveal
gas giant planets \citep{bond-moa53,ogle71}, only OGLE-2006-BLG-109 and
MOA-2007-BLG-400 \citep{dong-moa400} are highly sensitive to multiple planets
of Jupiter mass or less. The fact that OGLE-2006-BLG-109 did reveal a Jupiter-Saturn-like
system suggests that systems like our own are common among stars hosting gas giants
near or beyond the snow line.

The OGLE-2006-BLG-109L system is the first multi-planet system found by microlensing
but several other aspects of this event are also unique. The light curve reveals the
microlensing parallax effect, which yields a geometric 
mass measurement of the planetary
host star and both of the planets. 
In addition, the host star is five times brighter
than the source in the $H$ band, and it is detected in Keck AO images. The 
measured $H$-band brightness confirms the microlensing parallax mass 
measurement. 

In addition, the light curve shape is sensitive to the relative positions of the
planet at the pico-arcsecond level, and the light curve signal of the 
Saturn-mass planet is visible for 11 days. As a result, the orbital velocity of
this planet in the plane of the sky must be included to model the light
curve. Moreover, these effects also provide a weak constraint on the
orbital acceleration.

In this paper, we present some details of the analysis that was 
summarized by \citet{gaudi-ogle109}, and a 
new analysis of OGLE-2006-BLG-109 that includes the 
orbital motion of the OGLE-2006-BLG-109L system. With the assumption of
co-planar planetary orbits and orbital stability constraints, we derive limits on
the full set of orbital parameters for this Saturn-mass planet.

This paper is organized as follows. In \S~\ref{sec-data} we discuss the
image data and the photometric reductions. The modeling of the
light curve is described in \S~\ref{sec-model}, and a new method for
determining the angular radii of the source star and the Einstein Ring
is presented in \S~\ref{sec-radius}. The follow-up observations
that identify the planetary host star and their analysis are discussed in
\S~\ref{sec-follow}, and \S~\ref{sec-alt} discusses possible alternative
models for the light curve. The next section, \S~\ref{sec-orbits}, 
discusses the orbital motion modeling
and the conversion from measured fit parameters to inferred orbital 
parameters, and the Bayesian analysis used to find the constraints
on the physical orbital parameters is discussed in \S~\ref{sec-param}.
Limits on additional planets in the system are discussed in \S~\ref{sec-limits},
and we conclude in \S~\ref{sec-conclude}.

\section{Light Curve Data and Photometry}
\label{sec-data}

Microlensing event OGLE-2006-BLG-109 
(right ascension $\alpha = 17\,{\rm h}\,52\,{\rm m}\,34.51\,$sec,
declination $\delta =-30^\circ 05^\prime 16.0^{\prime\prime}$,
J2000) was announced by the 
Optical Gravitational Lensing Experiment (OGLE)
Collaboration Early Warning System (EWS)
\citep{ogle-ews} on 2006 March 26
(heliocentric Julian day $\sim 3821$), triggering follow-up observations by the
Microlensing Observations in Astrophysics
(MOA) collaboration using the 0.61m telescope at Mt. John Observatory in 
New Zealand. Two days later, the OGLE Early-Early Warning System
(EEWS) \citep{ogle-pipeline}
detected a deviation from the standard single lens
light curve. This led OGLE to take three additional images of the field
that includes this event on the same night. These additional images
confirmed the deviation, and so OGLE issued an anomaly alert. The
OGLE alert message noted that ``short-lived, low amplitude" 
anomalies can be caused by planetary companions, and implied
that there was a good chance that this was planetary signal.
This suggestion would prove to be correct, although in this case, the
planets orbiting the lens star would generate additional 
large amplitude anomalies over the subsequent 10 days.

This anomaly alert prompted follow-up observations from the
Microlensing Follow-Up Network ($\mu$FUN), 
Probing Lensing Anomalies Network (PLANET) and Robonet, although 
relatively few PLANET and
Robonet telescopes were available for bulge observations so early
in the season. Data were obtained from 11 different telescopes
spanning the globe. The two telescopes providing data from Chile
were the OGLE Warsaw 1.3m ($I$ band) and the Cerro Tololo
Inter-American Observatory (CTIO) Small and Moderate Aperture Research
Telescope System (SMARTS) 1.3m ($V$, $I$ and 
$H$ bands) telescopes. From the Southwestern US, data were provided through
the $\mu$FUN Collaboration from the
MDM 2.4m ($I$ band) and the Mt.~Lemmon 1.0m ($I$ band) telescopes in
Arizona, as well as the Aero8 0.3m (unfiltered) telescope in New Mexico, operated
by the Campo Catino Astronomical Observatory. Data from New Zealand came
from the MOA 0.61m telescope ($I$ band) at the Mt.~John Observatory 
on the South Island, and the Auckland 0.35m and Farm Cove 0.25m telescopes,
which are both $\mu$FUN telescopes located in the vicinity of Auckland.
Slightly further west is the PLANET Canopus 1.0m telescope 
($I$ band) in Tasmania, Australia.

The most sparsely covered longitudes are those west of Australia, and east
of Chile, where only Northern Hemisphere telescopes were available to
observe this event. The two telescopes filling this gap were the 
1.0m Wise telescope (unfiltered) in Israel, associated with $\mu$FUN, and
the Robonet/Liverpool 2.0m telescope ($R$ band) at La Palma in the
Canary Islands.

Although it is very difficult to get complete light curve coverage with these
telescopes so early in the Galactic bulge observing season, we were
able to obtain good coverage of four of the five caustic crossing
or cusp approach features in the light curve. Much of the first
feature was missed before the light curve anomaly was 
detected. Three of the remaining four features were visible
from New Zealand, where the availability of telescopes on both the 
North and South Islands allowed good coverage of these three features
even though each observatory missed one anomaly due to bad
weather.

An additional anomaly that would prove to be caused by a second
planet was observed from the Wise observatory in Israel and by
OGLE and $\mu$FUN-CTIO in Chile.

The photometry of the OGLE data and most of the $\mu$FUN data
were re-processed after the event was over using the OGLE
pipeline \citep{ogle-pipeline}. The exceptions were the 
CTIO $H$-band data, which were reduced with DOPHOT
\citep{dophot}, and the Wise data, which were reduced using
SoDOPHOT \citep{bennett-sod}. The MOA photometry was
done with the MOA difference imaging pipeline
\citep{bond-dia}, and the PLANET-Canopus 
and RoboNET-Liverpool Telescope data were reduced with
Pysis \citep{albrow-pysis}.

The color dependence of atmospheric extinction
can significantly affect the unfiltered (or ``clear" filter) photometry 
taken by some of the smaller telescopes due to the variation of
the airmass toward the target throughout the night
\citep{dong_ogle71}. This color dependence of the extinction
causes apparent variation in the relative photometry of stars
of different colors as the airmass changes. Since DIA photometry
is effectively normalized to some average of the brighter stars in
the field, this can introduce systematic photometry errors.
This can be corrected
by ensuring that the photometry of the target star is normalized to stars
with the same color as the source star. The ``clear"
filter data sets that cover important parts of the light curve
are the Auckland, Farm Cove, and Wise data sets. For these
data sets, we have determined the $V-I$ colors of all the 
bright stars in the images by matching the stars found in
SoDOPHOT reductions of  the clear filter data
sets to $V$ and $I$-band images from CTIO. 
The light curves of these stars were then generated
using the OGLE pipeline for the Auckland and Farm Cove
data and SoDOPHOT for the Wise data. A weighted mean of
these bright star  light curves with the same average color as
the target star was then used to normalize each 
OGLE-2006-BLG-109 light curve.

This normalization procedure was complicated by the fact
that the OGLE-2006-BLG-109 source star is blended with
a relatively bright clump giant star, as discussed in \S~\ref{sec-follow},
which is slightly redder than the source star. The separation between
the source and this blended star is only $0.35^{\prime\prime}$, so these
stars are never resolved in these clear filter images. Thus, the photometry
is being done on a blended target that changes color as the
lensing magnification changes.  This complicates the normalization
scheme for the unfiltered photometry as described above, because the
color of the blended image of the two stars is changing with time.
However, we expect this effect to be small because the difference
between the source and blend colors is only $\sim\Delta(V-I)=0.2$.
Nevertheless, we estimate the color
of the combined source-plus-blend image based on an early light
curve model and find that the correction is indeed small.
Of course, the exact values of the correction do depend somewhat
on the arbitrary choice of which model is used to make the
correction but, in fact, the correction is so small that any
reasonable model will give the same result to much better
than one-tenth of the estimated errors.

We have followed the standard microlensing analysis procedure of
rescaling the error bars in order to obtain $\chi^2/{\rm dof}$ for each
data set. Each data point had an assumed systematic error of 0.3\% 
added in quadrature to the error reported by the photometry code, and 
then the error bars were rescaled by the following factors:
1.5 for OGLE, 1.42 for MOA, 1.31 for CTIO $I$ band, 1.14 for
CTIO $H$ band, 1.55 for MDM, 0.52 for Mt.\ Lemmon, 1.93 for
Canopus, 2.3 for the Liverpool telescope, 1.8 for Auckland, 1.04 for 
Farm Cove, 1.5 for Wise, and 1.0 for Campo Catino.

The light curve modeling described in \S~\ref{sec-model} requires
limb darkening parameters because finite source effects are
important. The analysis of the extinction and color of the source
in \S~\ref{sec-radius} implies that the source has an approximate
spectral type of G0, with $T_{\rm eff} \approx 6000\,$K, and the
radius estimate implies that $\log g \approx 4.5$. From
\citet{claret2000}, this implies linear limb darkening parameters
of $u_V = 0.6630$, $u_R = 0.5887$, $u_I = 0.5090$, and
$u_H = 0.3292$. For the unfiltered Farm Cove and Auckland
data, $u_{FC} = 0.5413$, and
$u_{\rm Auck} = 0.5490$ are the preferred values. These were estimated
from color transformations derived using the method of
\citet{gould_col10}. The method indicates that the unfiltered or ``clear"
Farm Cove and Auckland passbands can be represented as
$C_{\rm FC} = 0.79I + 0.21V$ and $C_{\rm Auck} = 0.74I + 0.26V$,
and we have estimated the linear limb darkening coefficients with these
same linear transformations using the $I$ and $V$ band linear 
coefficients from \citet{claret2000}. We used 
$u_R$ for the Wise data set. Most of the modeling was
actually done with an earlier estimate of the source temperature,
$T_{\rm eff} \approx 6250\,$K and an older compilation of the
limb darkening parameters \citep{diaz95,claret95}. The values
used for most of the modeling runs were
$u_V = 0.633$, $u_R = 0.535$, $u_I = 0.456$, $u_H = 0.275$,
$u_{FC} = 0.502$, and $u_{\rm Auck} = 0.493$. The newer
limb darkening parameters yield a $\chi^2$ improvement of
$\Delta\chi^2 = 0.61$ and have no significant effect on any of the
model or derived parameters.

\section{Light Curve Modeling}
\label{sec-model}

OGLE-2006-BLG-109 is the first double-planet microlensing event and
the most complicated microlensing event yet to
be successfully modeled. The strongly magnified portion of the
light curve is shown in Fig.~\ref{fig-lc1}, with the peak region of the light curve
shown in Fig.~\ref{fig-lc2} and close-ups of the five caustic-crossing and
shown in Fig.~\ref{fig-lc3}. The light curve models shown in these figures
have 15 nonlinear parameters plus 24 linear parameters that describe
the source flux and background flux in each of the 12 instrumental passbands.
These linear parameters can be solved for exactly once the other 15 parameters
are specified, but determination of these 15 nonlinear parameters is
highly non-trivial.

Two of these 15 parameters describe microlensing parallax
\citep{refsdal-par,gould-par1,macho-par1} and three more describe the orbital motion
of one of the planets. The parallax parameters generate small 
perturbations in the light curve over long timescales, as well as 
very small perturbations near
peak magnification \citep{hardy95,holz_wald}, and the orbital motion parameters
provide relatively small perturbations to the timing and shape of the
various light curve features. So, it is sensible to initially ignore these
parameters, and to try and find an approximate solution with the
remaining 10 parameters. These ten parameters include the three
parameters of single lens events, the Einstein radius crossing time,
$t_E$, the separation of closest approach between the source and
the lens system center of mass, $u_0$, and the time of this
closest approach, $t_0$. An additional parameter, the source radius
crossing time, $t_\ast$, is needed for a small fraction of single lens
events and most planetary system lens events to describe the finite
source effects. There are 6 additional parameters that describe the
static configuration. The mass fractions of the two planets are
$\epsilon_1$ and $\epsilon_2$, and $\epsilon_3 = 1-\epsilon_1-\epsilon_2$
is the mass fraction of the planetary host star.
The angle between the source trajectory and the line connecting 
mass-1 to the center of mass of masses 2 and 3 is $\theta_{1\rm cm}$,
and the distance between mass-1 and the mass 2+3 center of mass
is $d_{1\rm cm}$. The remaining parameters are the distance
between masses 2 and 3, $d_{23}$ and the angle between the line connecting
these two masses with respect to the line between mass-1 and
the mass 2+3 center of mass, $\phi_{23}$. Both  $d_{1\rm cm}$ and $d_{23}$
are measured in Einstein radius units.

There are two basic aspects involved in modeling planetary microlensing 
events: the method used to calculate the microlensing light curve, and the
strategy employed to locate the appropriate model in the multi-dimensional
parameter space. The light curve calculations were done using variations
of the method first introduced by \citet{em_planet}, who developed the
first general binary-lens, finite source light curve calculation code and 
used it 
to demonstrate the sensitivity of the microlensing method to
low-mass planets. Their method uses the
point-source approximation except when the source is in the vicinity of
a caustic curve, where an inverse ray-shooting integration scheme is 
used. The point-source calculations for triple lens models 
are done using the method 
of \citet{rhie_triple}. At a preliminary stage of the calculation, one of 
the sub-groups doing the modeling also used the binary lens
superposition approximation 
\citep{rhie_pl96,bozza99,ratt,han-bin_sup,kubas-ogle390}.

The search for solutions for triple-lens systems is substantially more difficult
than for binary-lens models due to the additional three parameters needed
to describe the second planet.
The method of  \citet{bennett-himag} is particularly 
well suited to such events. It uses
a grid only for the initial conditions, and then allows all parameters to 
vary while minimizing $\chi^2$ from these initial points. This allows the
solution to be found in a fully automated manner without a huge increase
in the computation time due to the larger dimension of parameter space.
As we shall see, this method is also relatively efficient for modeling events
that include lens orbital motion.
For this event, however,  the modeling effort is aided by the fact that 
the major features in the light curve
are covered reasonably well and by the fact that the features due to the 
different planets do not overlap on the light curve. So, the global
parameter search part of the  \citet{bennett-himag} method was not used.

The light curve modeling began while the event was still in progress, and
modeling done after the detection of features 1, 2, and 3 shown in
Figs.~\ref{fig-lc1}-\ref{fig-lc3} was able to predict the future
behavior of the event, and in particular feature 5
\citep{gaudi-ogle109}.  However, this model did not predict feature 4.
Because the light curve
deviations due to multiple planets often resemble the superposition of
the planetary deviations due to each planet alone
\citep{rhie_pl96,bozza99,ratt,han-bin_sup},
this was considered a strong hint that the signal of a second planet
was present. Nevertheless, two subgroups proceeded with systematic 
attempts to model this event with a single planet model using
methods similar to that of \citet{cassan-meth}, but developed 
independently. Attempts were made to find binary lens models 
that could explain 4 of the 5 features, and it was found that this
was only possible when the fourth feature was excluded, just as
the preliminary analysis during the event had indicated.

The binary superposition approximation \citep{han-bin_sup} with Markov
Chain Monte Carlo (MCMC) minimization was then used to search for
approximate static, two-planet solutions.  This search succeeded in
identifying classes of models that could explain the basic features of
the light curve.  A second group then used this class of models as an
approximate input to a full triple lens modeling code
\citep{rhie_triple,mps-97blg41} and the light curve calculation
and \citet{metrop} $\chi^2$ minimization recipe of
\citet{bennett-himag}. The 
initial fitting using static, triple-lens models
could almost explain the OGLE-2006-BLG-109 light curve, but it proved
impossible to fit features 1 and 5 with the same static lens model.

However, the typical orbital motion of a planet in a microlensing event
is about $10^{-3}R_E$ per day, and for an event like OGLE-2006-BLG-109,
for which the main signal comes from a planet near the Einstein Ring, the
radial motion of the caustic has an amplitude similar to, but slightly 
larger than the motion of the planet. (The angular motion of the caustic
is much smaller than this.) As a result, over the 11 days between features
1 and 5, we can expect that the radial position of the caustic curve
will change by $\sim 0.01 R_E$, which is more than enough to have
a significant effect on the light curve. So, it was obvious that the orbital
motion of the planet close the Einstein Ring would have a significant
effect on the light curve, and the failure of the static, triple-lens models
indicated that orbital motion must be included.

The inclusion of orbital motion adds a significant complication to the
modeling of microlensing events. For most high magnification 
microlensing events, it is possible to
significantly reduce the light curve computation time by making use of
the fact that the lens configuration stays approximately the same
throughout the event.  With this static approximation, it is possible to
make and store a single map of the two-dimensional magnification of
the lens as a function of source position, from which one can quickly
and efficiently draw many trial one-dimensional light curves, thus
reusing the information for many different observations.
This is the basis of the inverse ray-shooting \citep{wamb,ratt}
and magnification map \citep{dong-ogle343} methods,
which densely ray-shoot
broad swaths of the image plane to determine the magnification of a large
number of source positions. In particular, the calculations of high
magnification events light curves can be made substantially more
efficient in this way by only ray-shooting the images near the
Einstein Ring.  Unfortunately, these shortcuts cannot be used for
modeling an event with orbital motion because the lens configuration
is different for every observation.

The \citet{em_planet} method also uses a version of the ray-shooting
method for its finite source calculations, but rather than sampling
the entire image plane, simply samples those images that are created
for a given source position. The current version of this method
\citep{bennett-himag} allows the option to store the rays shot in
the vicinity of the Einstein Ring, so that they do not have to be 
re-calculated for different observations, but this
optimization cannot be used for this event. This method also
employs some numerical integration improvements that 
speed up these high magnification event calculations by a
large factor ($\sim 100$). So, despite the fact that the
orbital motion prevents the ray-storage optimization, the
method of \citet{bennett-himag}
has been proved efficient enough to model this event.

The importance of orbital motion can be seen in Figure~\ref{fig-caustic},
which shows a time series of caustic positions at intervals of 
$2.9\,$days. The red curve is the caustic position at the time
of feature 1. It reveals the two caustic crossing seen in the 
light curve model. The second caustic entrance (feature 2)
occurs about a
day before the time of the green caustic curve, and the highest
magnification peak of the light curve occurs on the caustic exit
(feature 3)
at the time of the black caustic curve. The final cusp  approach 
(feature 5) occurs at the time of the blue caustic curve.

It is apparent that the radial motion of the caustic in 
Figure~\ref{fig-caustic} is much larger than the rotation
of the caustic. This is, in fact, a general feature of orbital
motion in high magnification microlensing events. This orbital motion
is most easily detected when a planet is close to the Einstein 
ring, so that the caustic is extended and the planetary signal
has a long duration. But in this situation, the radial motion of
the central caustic has a velocity similar to the velocity of the
planet. However, the angular motion of the caustic is smaller than
the planet's angular velocity by the distance of the caustic from
the origin (center of mass) in Einstein radius units. For this lens
system, the caustic extends to a distance of $\sim 0.15$ Einstein
radii from the center of mass, but the source does not encounter the
caustic at this furthest point. 
With the actual source trajectory, the caustic is encountered at a
distance of $\sim 0.05$ Einstein radii, and so the 
angular motion of the caustic is suppressed by a factor of 20
compared to the radial motion. This situation is typical, so we should generally
expect that the effect of the angular motion of the planet in the plane
of the sky will usually be compressed by a factor of $\sim 10$ compared
to the apparent radial motion. Thus, it will usually not be useful
to approximate orbital motion as rotation in the plane of the sky
for high magnification events, although this is much more efficient
to investigate computationally \citep{ratt}.

The basic procedure used to find the solution for this event has been to
proceed from a simple single-planet plus star model without orbital motion,
and then to add additional effects one-by-one. This is done in the 
following order:
\begin{enumerate}
\item A static model with a star and a single planet, approximately
matching all features, except feature 4.
\item Add a second planet to account for feature 4. The best
static two-planet model cannot simultaneously explain the
details of features 1 and 5.
\item Include orbital motion of the Saturn-mass planet, which is responsible
for features 1-3 and 5.
\item Add microlensing parallax to include the
effect of the orbital motion of the Earth.
\end{enumerate}

The orbital motion of the Saturn-like planet, like that of any other mass, can
be described by 7 parameters. These include 6 parameters,
such as a three-dimensionalimensional position and velocity, that can describe the 
initial conditions, plus the mass of the system, which is required to determine
the future orbital motion.
For an object in a Keplerian orbit, the conventional parameters
include parameters like eccentricity and time of periapsis, which are not
likely to be well constrained by a microlensing light curve, when the planetary
signal only lasts 11 days. So, we have selected a set of parameters that
can be separated according to how well they are constrained by the
microlensing light curve. The two-dimensional position of the planet
in the plane of the sky is described by two of the static lens-system
parameters. The lowest order effect of orbital motion is simply the
two-dimensional velocity of the planet in the plane of the sky
described by $\dot d_{23x}$ and $\dot d_{23y}$ (where the $x$-direction
is defined by the line from mass 1 to the mass 2-3 center of mass). 
For the models we consider here, the Jupiter-analog planet is mass
1, the Saturn analog planet is mass 2, and the host star is mass 3,
so it is masses 2 and 3 that are in orbital motion.

The two-dimensional position and velocity of the Saturn-analog
planet (with respect to the host star) gives us
4 parameters (out of 7 total) that are tightly constrained. The
simplest  possibility would simply to use a constant
velocity for the Saturn-mass planet, and ignore any higher-order terms
such as the acceleration of the planet.  Although this
approximation does not correspond to a consistent orbital solution, it
is reasonable for events for which orbital motion is only weakly
detected, where the higher-order terms are not constrained
\citep{dong_ogle71}.  However, there are a of couple problems with
adopting this approach.  One danger of such a strategy
is that this unphysical motion could have a significant effect
elsewhere in the light curve.  For example, the planetary caustic
could move close enough to the source trajectory to be detected. While
it is unlikely that this would happen for the best fit model, but it
is more likely to happen for somewhat disfavored models that may be
explored when we determine the parameter uncertainties with long
Markov Chain Monte Carlo (MCMC) runs.

Another disadvantage of this unphysical parameterization scheme
is that it is possible that other parameters, beyond the two-dimensional
velocities, will be constrained by the light curve data, and in fact, we find that 
this is indeed the case for OGLE-2006-BLG-109.  However, in order to minimize
the number of unconstrained parameters, we do not permit the full freedom
that orbital motion allows. Instead, we restrict the orbital motion to circular
orbits and add a single additional parameter, the orbital period, $T_{\rm orb}$.

The next light curve feature to be included is microlensing parallax. 
This can be described \citep{gould-par00,exoplanet_book} by the 
projected Einstein radius, 
$\repbold$, which has a magnitude, $\rep$, equal to the Einstein radius 
projected (from the source) to the position of the observer and a direction
parallel to the lens-source relative proper motion. This can be measured
in a variety of ways. The most common way to measure $\rep$ is through
the effect of the orbital motion of the Earth \citep{gould-par1,macho-par1},
but it has also been measured via the spatial separation of 
different telescopes on the Earth \citep{hardy95,holz_wald,ogle224} and via
observations from satellites in heliocentric orbit \citep{dong-ogle05smc1}.
For OGLE-2006-BLG-109, we find that the orbital parallax effect is
dominant, as the event is quite long with $t_E > 120\,$days, but we
also see a significant terrestrial parallax effect, due to the different locations
of the observatories on the Earth. However, this terrestrial parallax was not
included in the initial modeling reported in \citet{gaudi-ogle109}.

While it is most convenient to describe microlensing parallax in terms of
the projected Einstein radius vector, $\repbold$, this is not a convenient
parameter to use for microlensing light curve fits. It is often the case 
that the microlensing parallax signal is weak, but this implies that
$\rep \rightarrow \infty$. So, it is conventional to use the 
microlensing parallax vector, $\piEbold \parallel \repbold$, which has a 
magnitude, $\pi_E = 1\,{\rm AU}/\rep$. As advocated by \citet{gould-jerkpar},
we work in a geocentric frame, which is at rest with respect to the Earth
at a fixed time (${\rm HJD} = 2,453,831\,$days in this case). We use
polar coordinates, $\pi_E$ and $\phi_E$ to describe $\piEbold$, so that 
the north and east components of $\piEbold$ are
$\pi_{E,N} = \pi_E \cos\phi_E$ and $\pi_{E,E} = \pi_E \sin\phi_E$.

The addition of microlensing parallax yields 15 nonlinear parameters
needed to model this event. In addition, there are two linear parameters
for each telescope and passband to describe the source flux and blended
(or background) flux for each of these data sets. These linear parameters
are solved for exactly for each set of nonlinear parameters that is
considered.

With these 15 parameters, we can now find
a fit that explains all the features of the
OGLE-2006-BLG-109 light curve, but this is not enough to ensure that we
have found the best solution. In particular, the planetary orbital motion and
the microlensing parallax can change the relative motion of the source
and the caustic curves in a similar way. Thus, it is possible that our
minimization routine and our MCMC runs will not fully explore the
degeneracy in these parameters. Therefore, we have done a series of
$\chi^2$ minimization runs with fixed values for $\pi_E$ ranging from
$\pi_E = 0.216$ to $\pi_E = 0.515$, which map out the trade-off between
parallax and orbital velocity and ensure that we have found the 
true $\chi^2$ minimum.

For single lens and binary lens events, there is usually an approximate
four-fold degeneracy in the microlensing parallax solutions
\citep{smith_par_acc,gould-jerkpar}, although this degeneracy can
be broken for very long events or events with a significant
terrestrial parallax signal. One of these degeneracies is completely 
broken for triple lens events as it is related to the reflection symmetry
about the lens axis that exists for binary lens events. The third mass
breaks this reflection symmetry, so that only one approximate symmetry
remains. This remaining symmetry involves reflecting all of the lens system
parameters with respect to the parallax parameters, and then
shifting the parallax parameters by an amount that 
depends on the time of year when the event occurs. (In terms of 
the model parameters, this symmetry implies $u_0 \rightarrow -u_0$, 
$\theta_{1\rm cm} \rightarrow -\theta_{1\rm cm}$, and
$\phi_{23} \rightarrow -\phi{23}$.)
With the inclusion of terrestrial parallax, we find that this degenerate
parallax solution is disfavored by $\Delta\chi^2 = 37.9$, so that it is
not a viable solution.

For high magnification events, there is often also a
degeneracy associated with the $d \rightarrow 1/d$ transformation that
occurs when the source only probes the ``central caustic'' created by
a planet, whose shape is approximately invariant under this
transformation\citep{griest_saf}.  This degeneracy does not apply to the
Saturn-mass planet because it is close to the Einstein Ring. The 
source is observed while crossing some of the 
``planetary parts" of the caustic curve, which is not invariant under this 
transformation. However, this degeneracy does
apply to the Jupiter-mass planet, which is not so close to the Einstein
ring. In \citet{gaudi-ogle109}, we argued that the $d_J < 1$ solution
was disfavored by $\Delta\chi^2 = 11.4$ over the $d_J > 1$ solution.
However, the inclusion of terrestrial parallax improves the $\chi^2$ of the 
best $d_J < 1$ solution by 12.0 and the $\chi^2$ of the best $d_J > 1$
solution by 22.0, So, the $d_J > 1$ solutions do now seem nearly as
good as the $d_J < 1$ solutions. However, as we will see in 
\S~\ref{sec-orbits}, the parameters of these $d_J > 1$ solutions
do not generally correspond to planets with stable, co-planar orbits.
As a result, the $d_J < 1$ solutions remain strongly favored.


\section{Source Star Radius and Angular Einstein Radius}
\label{sec-radius}

The angular radius of the source star, $\theta_\ast$, is an important parameter, 
because we can combine it with the Einstein radius crossing time and
the source radius crossing time to yield the angular Einstein radius,
\begin{equation}
\theta_E = {\theta_* t_E\over t_*} \ ,
\label{eq-thetaE}
\end{equation}
which is needed to determine the masses of the lens star and its planets.

The angular radius or the source is routinely estimated from $V$- and
$I$-band data \citep{yoo_rad}, but the availability of $H$-band photometry
that is precise enough to yield an accurate $H$-band magnitude for the
source allows a much more precise method that uses three-color
$VIH$ photometry. This method is more precise because it allows
us to take into account slight spatial variations in the extinction and differing
extinction laws toward different lines of sight, and because it allows
us to use the more precise optical-IR 
surface brightness relations \citep{kervella_dwarf}. The $V$-$H$ relation
has a precision of $1.1\,$\%, to be compared to the 
nonlinear $V$-$I$ relations, which has
a precision of about $5\,$\% \citep{kervella_fouque}.

The first step toward estimating the intrinsic source star color and
magnitude is to identify the centroid of the red clump star distribution
in color-magnitude diagrams (CMDs) made of stars 
in the vicinity of the target star and then to compare these colors and
magnitudes to the known properties of these red clump giant stars.
We take the absolute red clump star magnitudes for the local
stellar population, with Hipparcos parallaxes, 
from \citet{alves_lmc_RC_dist}:
$M_K = -1.60 \pm 0.03$,  $M_I = -0.26 \pm 0.03$, and
$M_V = 0.73 \pm 0.03$. 
However, while red clump stars are approximate standard candles, their 
properties are known to vary with age, metallicity, and chemical 
composition. \citet{clump_agemet2} have calculated these corrections
to the red clump magnitudes for the known age and metallicity of 
giant stars in the Galactic bulge. They present two sets of Galactic
bulge corrections, one assuming a standard solar chemical
abundance distribution and another for $\alpha$-enhanced
abundances, which is expected to be more appropriate for the
Galactic bulge. We use the $\alpha$-enhanced corrections,
but we increase the error bars in proportion to the difference between
the $\alpha$-enhanced and solar metallicity corrections to 
account for uncertainties in  this correction and the possibility
of spatial dependence in the chemical composition of 
bulge red clump stars. This gives the following estimates
of the absolute magnitude of the red clump:
\begin{eqnarray}
M_{Krc} =& -1.49 \pm 0.03 \\
\label{eq_rc_amagH}
M_{Hrc} =& -1.41 \pm 0.04 \\
\label{eq_rc_amagI}
M_{Irc} =& -0.25 \pm 0.05 \\
M_{Vrc} =& 0.79 \pm 0.08 \ ,
\label{eq_rc_amagV}
\end{eqnarray}
where we have used the \citet{bessell_brett} giant star
color-color relations to derive the $H$-band magnitude.

These magnitudes must now be compared to the observed
red clump magnitudes in the vicinity of the OGLE-2006-BLG-109S
source star. Figures~\ref{fig-cmd_VI}-\ref{fig-cmd_IH} show 
CMDs of all the stars within $1^\prime$
of OGLE-2006-BLG-109S. Figure~\ref{fig-cmd_VI} uses $V$ and
$I$ magnitudes from both OGLE-II and CTIO. The CTIO magnitudes
are only shown for stars that are identified in all three bands observed
with the CTIO-ANDICAM instrument: $V$, $I$, and $H$. Since the $H$-band
images are relatively shallow, the 3-band
CTIO photometry (shown as blue dots in
Figure~\ref{fig-cmd_VI}) does not extend to stars much fainter
than the red clump. The small black dots indicate the OGLE-II
photometry, which is not constrained to have stellar counterpart
identified in the $H$-band data. The source star and the centroid
of the red clump star distribution are indicated by the large 
black and red dots, respectively, and the location of the bright star
that is $0.35^{\prime\prime}$ from the source is indicated by the
large green dot.

The $V$ and $I$ magnitudes from CTIO have been calibrated to
the same OGLE-II system used for the OGLE data \citep{ogle2_BVImap},
and the CTIO $H$-band data have been calibrated to 2MASS 
\citep{2mass_cal}\footnote{Improved calibrations are available at 
\url{http://www.ipac.caltech.edu/2mass/releases/allsky/doc/sec6\_4b.html}}
using a set of common stars that have been found to have no close 
neighbors in the higher angular resolution CTIO frames.

We locate the centroids of the red clump distribution in these
CMDs by first creating a smoothed stellar density distribution
by convolving each CMD with a two-dimensional Gaussian
with $\sigma = 0.1$ magnitudes. We then find the maximum
of the red clump distribution in this smoothed CMD. This was
done for each of the CMDs plotted in 
Figures~\ref{fig-cmd_VI}-\ref{fig-cmd_IH}, including both the
CTIO and OGLE-II $V$-$I$ CMDs shown in Figure~\ref{fig-cmd_VI}.

We also make a correction to the clump position measured
in the field of OGLE-2006-BLG-109S due to crowding. The
need for such a correction is apparent from artificial star
tests of photometry in crowded stellar fields \citep{macho-lmc5.7eff}.
The photometry codes used to make the magnitude measurements
for the CMDs are less complete in finding faint stars that are 
located under the point-spread functions (PSFs)  of brighter stars. 
So, these faint stars are more likely to be detected when they are 
not in the vicinity of brighter stars. Thus, the unresolved star 
background is, on average brighter under the PSF's of bright stars.
Since these photometry codes do not account for this, they tend
to slightly overestimate the brightnesses of these brighter stars.
The appropriate correction can be estimated from the OGLE-II
analysis of microlensed red clump stars by \citet{ogle2_blg_tau}. 
The source brightnesses for these are
determined by the microlens model fit, and so they are 
independent of this photometric bias. Considering only the 29 stars
from \citet{ogle2_blg_tau} with source flux fractions $f_s > 0.5$, we
find $\VEV{f_s} = 0.95 \pm 0.03$. (Stars with $f_s > 0.5$ are
generally blends of two giant stars and do not appear to be part
of the clump.)
This is a weighted average, but we
have added an error of 0.1 in quadrature to the fit uncertainty reported
by \citet{ogle2_blg_tau} to prevent the shot noise
from the few events with very small
$f_s$ fit uncertainties from increasing the uncertainties. This correction
might have some color dependence as the red clump stars are
likely to have a color that is somewhat different than the stars
responsible for this blending effect. However, we only have an
estimate of this effect in the $I$-band, so we
add 0.05 mag correction to the $VIH$ magnitude values
determined from the CMDs. The errors in this procedure are
dominated by the uncertainty in locating the clump centroid
in the CMDs. We tried a variety of different smoothing radii
to find the centroids for all 4 CMDs shown in 
Figures~\ref{fig-cmd_VI}-\ref{fig-cmd_IH}. (Figure~\ref{fig-cmd_VI}
has CMDs from both OGLE-II and CTIO.) The clump centroids
for all CMDs with all reasonable smoothing radii are consistent
with these adopted red clump magnitudes and uncertainties,
\begin{eqnarray}
\label{eq_rc_magH}
H_{rc} =& 13.76 \pm 0.10 \\
\label{eq_rc_magI}
I_{rc} =& 16.28 \pm 0.10 \\
V_{rc} =& 18.88 \pm 0.10 \ ,
\label{eq_rc_magV}
\end{eqnarray}
for the red clump stars within $1^\prime$ of
OGLE-2006-BLG-109S. 

These measured red clump magnitudes allow a comparison of the
predicted absolute red clump magnitudes given in 
eqs.~\ref{eq_rc_amagH}-\ref{eq_rc_amagV} with the measured
magnitudes from eqs.~\ref{eq_rc_magH}-\ref{eq_rc_magV}. 
We now fit for the extinction assuming the \citet{cardelli_ext} 
extinction law,
and a Galactic center distance of $R_0 = 8.0\pm 0.2\,$kpc.
This model has two parameters, and there are three constraints,
the $VIH$ magnitudes of the red clump centroid. The best fit
yields $A_V = 3.47\pm 0.07$ and 
$R_{VI} = A_V/(A_V-A_I) = 2.41{+0.24\atop -0.19}$. 
These parameters
also imply that $A_I = 2.03\pm 0.08$, $A_H = 0.60\pm 0.04$
and $R_V = 2.96{+0.70\atop -0.46}$. With two model parameters
and three measurements, we have $\chi^2 = 0.151$ for a single
degree of freedom.
We note that the  \citet{cardelli_ext} law uses an infrared extinction
law that does not quite agree with direct measurements toward the
Galactic bulge \citep{nishiyama_ext}. However, since we are only
considering one infrared passband (the $H$ band), we are not 
forcing infrared extinction to follow the Cardelli law between different
infrared passbands. So, this deficiency
of the Cardelli law will have little influence on our results.

If we assume that the source star has the same extinction as the
average of the red clump stars within $1^\prime$ of the source,
then we can use the \citet{kervella_dwarf} ($V$-$H$,$H$) color-radius
relations to derive a source radius of 
$0.474{+0.020\atop -0.018}\,\mu$as. Note that these Kervella 
relations use infrared magnitudes that are effectively on the
\citet{bessell_brett} system, so we must convert from the 
Two Micron All Sky Survey (2MASS)
system to Bessell \& Brett using the formulae in \citet{2mass_cal}.

However, the extinction toward the source is not
identical to the average extinction toward the clump giants
within $1^\prime$ of its position, although experience with
high resolution spectra of microlensed bulge
main sequence stars indicates that the extinction toward the source
is not likely to differ from the average extinction toward the
neighboring red clump stars by more than 5\%
\citep{moa310_311_spec}. With measurements of the source
star brightness in the three $VIH$ passbands, we can make
use of the \citet{kenyon_hartmann} color-color relations for
dwarf stars to demand that the extinction is consistent with the
colors of a main sequence star, as well as approximately
matching the extinction law and value derived for the red clump
stars. We use
\begin{eqnarray}
\label{eq_s_magH}
H_s =& 18.876 \pm 0.030 \\
\label{eq_s_magI}
I_s =& 20.935 \pm 0.030 \\
V_s =& 23.110 \pm 0.030 \ ,
\label{eq_s_magV}
\end{eqnarray}
for the magnitudes of the source star. Note that we deliberately 
overestimate the error bars in eqs.~\ref{eq_s_magH}-\ref{eq_s_magV}
to account for the uncertainties in the color-color relations.
We also require that  
the \citet{cardelli_ext} extinction parameter, $R_v$, matches the
value determined for the red clump stars within $1^\prime$
of the source and that the total $H$-band extinction be
within 5\% of the value determined for the red clump
stars. Using the magnitudes in 
eqs.~\ref{eq_s_magH}-\ref{eq_s_magV} and the derived
extinction estimates, we then use the ($V$-$H$,$H$) color-radius
relation of \citet{kervella_dwarf} to derive the angular
radius of the source star.

We use two methods to impose these constraints and determine
the source radius. The first method is to impose these constraints
using a MCMC calculation that selects star colors at random with
the requirement that they obey the \citet{kenyon_hartmann} 
color-color relations and $R_v$ and $A_H$ selected from the
probability distributions mentioned above. The $\chi^2$ is calculated
for the $V-H$, $V-I$, and $I-H$ colors derived from
eqs.~\ref{eq_s_magH}-\ref{eq_s_magV}. This $\chi^2$ is multiplied
by $2/3$ before being used in the \citet{metrop} algorithm to
account for the fact that these colors satisfy
$V-H = (V-I) - (I-H)$. Since each point on the Markov Chain gives
unique values for the dereddened source magnitude, $H_{s0}$ and 
color, $(V-H)_{s0}$, each point on the chain specifies a unique
$\theta_\ast$ value.

Our second method is
to consider all combinations of $R_v$ and source color (subject to 
the color-color relations), and then adjust $A_H$ to minimize the 
$\chi^2$ with the colors constrained by the color-color relations. This
results in the $\chi^2$ versus $\theta_\ast$ distribution shown in 
Figure~\ref{fig-radius}, which is approximately bounded by the 
$\theta_\ast = 0.468 \pm 0.012\,\mu$as curve derived from the
MCMC calculation. However, this estimate does not include the
quoted 1.1\% uncertainty in the \citet{kervella_dwarf}
color-radius relation. When this is added, we find
\begin{equation}
\theta_\ast = 0.468\pm 0.013\,\mu{\rm as} ,
\label{eq-thetaS}
\end{equation}
for the angular radius of the source star. This can be combined with 
two light curve parameters, the
Einstein radius crossing time, $t_E$, and the source radius crossing
time, $t_*$, to yield the angular Einstein radius,
$\theta_E = \theta_* t_E/t_* = 1.505\,$mas, using the
best fit parameters listed in Table~\ref{tab-mparam}.
We do not include
error bars here because the complete analysis of the uncertainties
is given in \S~\ref{sec-param}.

With this determination of 
$\theta_E$, we can use the 
length of the microlensing parallax vector to determine the
lens mass, 
\begin{equation}
M_L = {\theta_E \rep c^2\over 4G}
    = {\theta_E c^2 {\rm AU}\over 4G\pi_E} 
    = {\theta_E \msun\over (8.1439\,{\rm mas}) \pi_E} \ .
\label{eq-mass}
\end{equation}
Using the best fit value of $\pi_E = 0.3619$ from Table~\ref{tab-mparam}
this gives $M_L = 0.51\,\msun$.
Since the distance to the source star is generally known, at
least approximately, we can also determine the distance to the
lens star:
\begin{equation}
D_L =  {D_S\over 1 + {\pi_E \theta_E D_S\over 1\,{\rm AU}}} \ .
\label{eq-Dl}
\end{equation}
If we assume $D_S = 8.0\,$kpc and the fit parameters from 
Table~\ref{tab-mparam}, this yields $D_L = 1.49\,$kpc. 
Note that for our adopted parameters, the fractional error
in $D_L$ induced by the error in $D_S$ is
a factor of 0.19 smaller than the
fractional error in $D_S$ itself, so the uncertainty in $D_L$ is quite small.

\section{Physical Constraints on Model Parameters}
\label{sec-constrain}

In principle, it is possible for a planet detected in a microlensing
event to be physically unrelated to the lens star, 
but this is quite unlikely. If the signal
occurs near the peak of a high magnification event, the probability
is extremely small, $\simlt 10^{-8}$ per event. So, with less than
100 high magnification microlensing events observed, the probability
that an unrelated planet has been seen is $< 10^{-6}$. (Furthermore,
if two lens objects are closely aligned by chance, they will 
generally be at different distances,
so the usual binary lens magnification
equations do not apply \citep{rhie_bennett09}.)
So, we will assume that the planets
detected in the OGLE-2006-BLG-109 light curve are in orbit about
the lens star, OGLE-2006-BLG-109S. With this assumption,
we can use the lens mass determined from eq.~\ref{eq-mass}
to constrain the parameters describing the orbital motion:
the velocity in the plane of the sky, described by
$\dot{d}_{23x}$ and $\dot{d}_{23y}$, and the orbital
period, $T_{\rm orb}$. 

When the mass of the lens system is unknown, there are 7 parameters
needed to describe the relative orbital motion of two masses.
(These can be taken to be the initial relative position and velocity,
plus the reduced mass that is needed for the gravitational equation
of motion.) Since we know the lens mass from eq.~\ref{eq-mass}, it
might seem that we can reduce the number of free parameters to 
6, but there is a complication. The lens separation and relative
velocity parameters are given in units of the (linear) Einstein Ring radius,
$R_E$, but the gravitational equation of motion requires standard
physical length units. So, in order to make use of the measured
lens system mass to reduce the number of parameters, we need to
know the linear Einstein radius $R_E = D_L \theta_E$. Although we can use
eq.~\ref{eq-Dl} to give us the lens distance, $D_S$ is only
approximately known, and therefore this cannot give a precise
constraint on the lens parameters.

We will consider two sets of constraints, corresponding to two
different interpretations of our set of orbital parameters. As mentioned in
\S~\ref{sec-model}, our models have
the three parameters describing orbital motion and two
parameters giving the instantaneous position of the second
planet ($d_{23}$ and $\phi_{23}$)
so our models have 5 parameters to describe the orbits. As discussed
in \S~\ref{sec-model}, these parameters are sufficient to characterize
a circular orbit, which is what our lensing model assumes. If we
restrict our consideration to circular orbits, then we can use
eqs.~\ref{eq-mass} and \ref{eq-Dl} to constrain the model parameters.
However, since $D_S$ is known only approximately, we do not impose
a hard constraint on the fit parameters. Instead, we invert 
eqs.~\ref{eq-mass} and \ref{eq-Dl} to find an expression for $D_S$.
We then add a term to $\chi^2$ of the form 
\begin{equation}
\chi^2_{D_S} = (D_S-D_{S0})^2/\sigma^2_{D_S} \ ,
\label{eq-chi2Dl}
\end{equation}
to impose a ``soft"
constraint on the lens parameters during the modeling process.
The models we present here 
used $D_{S0} = 8.0\,$kpc and $\sigma_{D_S} = 1.5\,$kpc.
This $\sigma_{D_S}$ is a little larger than the prediction of most bulge
models, but we pick a conservative value as we do not need a tight
circular orbit constraint.

The advantage of this constrained circular orbit modeling scheme is that 
it prevents the modeling runs from straying into the wide swaths of parameter
space that do not correspond to physical orbits. Also, for this event, the
orbital period parameter, $T_{\rm orb}$, is only very weakly constrained,
so there is little danger that this constraint will prevent the modeling and
MCMC runs from reaching the vicinity of the correct model in parameter
space. If $T_{\rm orb}$ were more tightly constrained, then it might be sensible
to try a weaker constraint on $D_{S}$
that just ensures that the fit velocity is not
large enough to make the total energy is positive, implying an
unbound system.

While the constrained circular orbit scheme is a useful way to explore 
parameter space to ensure that all the viable models are considered, it
cannot be used to work out the constraints on the physical orbital
parameters of the system. When we consider this question in \S~\ref{sec-orbits} and
\ref{sec-param}, we use a different interpretation of the $T_{\rm
orb}$ parameter.  Formally, $T_{\rm orb}$ is the period of the
circular orbit of the planet. However, if we expand the
motion of the Saturn-mass planetary lens in a Taylor series,
we will see that there is an alternative interpretation. The
first-order correction to the static two-dimensional 
Saturn-mass planet position
used for most events is the two-dimensional velocity of this
planet, which is strongly constrained by the photometric data for
OGLE-2006-BLG-109. However, the direction of the next order,
acceleration term is constrained by Newton's law of gravity to 
be toward the stellar lens mass. Since our orbit model has
only a single additional parameter beyond the first-order velocity
parameters, this additional parameter, $T_{\rm orb}$, must be
equivalent to the orbital acceleration to second order. And since
$T_{\rm orb}$ is weakly constrained by the data, we can be confident
that the higher order terms are essentially unconstrained by the data.
This allows us to make an alternative interpretation of the $T_{\rm orb}$ 
parameter for this event. We can interpret it as the orbital acceleration.
This interpretation is used in \S~\ref{sec-orbits} and \ref{sec-param}, 
where we derive constraints on the orbital parameters of
OGLE-2006-BLG-109Lc (the Saturn-mass planet).

\section{Follow-up Observations and Analysis}
\label{sec-follow}

We have obtained follow-up adaptive optics (AO) images from the
Keck Observatory and spectra from the Magellan Telescope
in an attempt to characterize the planetary host star and the neighboring
bright red clump star. The Keck observations and interpretation are
discussed in \S~\ref{sec-keck}-\ref{sec-lens_bcont}, and the Magellan
spectra are discussed in \S~\ref{sec-spec}.

\subsection{Keck Observations}
\label{sec-keck}

The close-up of the OGLE finding chart shown in the left-hand panel of
Figure~\ref{fig-keck} indicates that the event appears to be centered
on a relatively bright star. This is the red clump giant star indicated
by the green dots in the CMDs shown in 
Figures~\ref{fig-cmd_VI}-\ref{fig-cmd_IH}. These color-magnitude
diagrams also indicate that the source is some 4.1-4.6 mag
fainter than this red clump star (depending on the passband),
so this bright star is obviously not the source. We
initially considered the possibility that this could be the lens star,
which would have been inconsistent with the photometric properties
implied by the lens mass and distance implied by the 
microlensing parallax and $\theta_E$ measurement as 
discussed in \S~\ref{sec-radius}.
However, an astrometric analysis of the OGLE data indicated
that the lensing event is centered $0.31^{\prime\prime}$ north
of the centroid of the relatively bright object at the center of
the crosshairs in the left panel of Figure~\ref{fig-keck}. This means
that most of the flux in this star-like image is due to a star that
is unrelated to the microlensing event. (We show that this star
does not produce any significant lensing effect in \S~\ref{sec-limits}.)

In 2007 July, we obtained $H$- and $K$-band AO
images of OGLE-2006-BLG-109 with the NIRC2 instrument on the 
Keck-2 telescope. The AO correction was made using
a natural guide star located $38^{\prime\prime}$ from the target. 
As a result of this relatively large angular distance to the guide star,
the AO correction is not as good as the Keck AO system normally
provides. The image FWHM values range from 
$0.09^{\prime\prime}$ to $0.13^{\prime\prime}$ in the 
$K$ band and from $0.15^{\prime\prime}$ to $0.25^{\prime\prime}$
in the $H$ band. The right-hand panel of Figure~\ref{fig-keck}
shows the best seeing $H$-band image. While this is not
a very good correction by Keck standards, it is fine for our
main purpose which is to resolve the combined image of the
lens and source stars 
(in the red circle in Figure~\ref{fig-keck}) from
this much brighter, unrelated star (labeled A in
Figure~\ref{fig-keck}).

Although the $K$-band images have better image quality, we
focus on the $H$-band data because we have $H$-band data
from CTIO during the event that allows us to determine the 
brightness of the source star in the $H$ band. The reduction of
the Keck AO data requires a crowded field photometry code because
of the significant overlap in the images of our target star with
the bright red clump star, labeled ``A" in Figure~\ref{fig-keck}.
The reduction is done with DAOPHOT \citep{allframe} using the
``penny1" PSF function with a 62 pixel radius. (The pixels
subtend $0.01^{\prime\prime}$.)

The PSF shape changes significantly across the AO images,
perhaps as a result of the relatively large $38^{\prime\prime}$
angular distance to the guide star and the high airmass of the
Galactic bulge as observed from Hawaii. In order to avoid 
significant photometry errors due to the spatial dependence of the 
PSF, we only use two relatively bright stars within $3^{\prime\prime}$
to do the relative calibration of the Keck $H$-band photometry to 
CTIO. These stars are labeled A and B in Figure~\ref{fig-keck}.
The CTIO photometry is then calibrated to the 2MASS 
system \citep{2mass} using 16 stars in common between CTIO and
2MASS that have been specifically selected to be uncrowded 
in the CTIO frames to avoid systematic errors due to blending
in the relatively poor seeing 2MASS images.

This analysis yields a total lens plus source magnitude of
$H_{ls} = 16.99 \pm 0.04$, which can be combined with the
source magnitude of $H_s = 18.876 \pm 0.014$ from the light curve
model to yield a lens magnitude of $H_L = 17.17 \pm 0.05$.
The contributions to this final uncertainty in $H_l$ are 1\% from
the CTIO-2MASS calibration, 2\% from the CTIO-Keck calibration
and 4\% from the Keck photometry.

We can now compare this magnitude to the expectations based
on our calculations in \S~\ref{sec-radius}. 
Using the mass-luminosity
relations of \citet{kroupa_tout}, we predict an absolute $H$ magnitude
of 5.94 for the $M_L = 0.51\,\msun$ lens star. At a distance of 
$D_L = 1.49\,$kpc, this yields a dereddened lens magnitude of
$H_{L0} = 16.81$. However, the ``best fit" model considered in
\S~\ref{sec-radius} actually included an $H$-band magnitude constraint,
so to avoid circular reasoning, we should actually use the best
fit model without such a constraint. This model is quite similar,
but predicts $M_L = 0.52\,\msun$ and $D_L = 1.53\,$kpc, which
would imply an absolute $H$ magnitude of 5.87, and $H_{L0} = 16.79$.
Now, the extinction in the foreground of the
source is $A_H = 0.60$ mag, but the source is located
at Galactic coordinates of $l = -0.2086^\circ$ and $b = -1.8901^\circ$,
which means that the lens system is about $35\,$pc south of the
Galactic plane. As a result, it is likely that there is 
a significant amount of dust in the foreground of the lens system
and between the lens and source. So, we might guess that the
extinction toward the lens is $A_{HL} \approx 0.3 \pm 0.2$. This leads
to the prediction of $H_L = 17.09\pm 0.20$, which of course, matches
our Keck measurement quite well.

\subsection{Must the Excess $H$-Band Flux Seen by Keck Be from the Lens Star?}
\label{sec-excess}

Before ending our discussion of the Keck observations, we should also
consider the possibility that another star besides the planet host star is
responsible for this detected $H$-band flux. There have now been
several events that have also had the detection of an additional
star superimposed on top of the source in a high resolution
image. For the first two planets found by microlensing, images
with the Hubble Space Telescope (HST) indicated 
different centroid positions for the source plus blend star in
different colors, which was consistent with the expected
offset due to the lens-source proper motion determined
from the light curve  \citep{bennett-moa53,dong_ogle71}. So, for
these events, it is quite likely that the additional flux is due to the
planetary host/lens star. However, for MOA-2008-BLG-310, 
where excess flux that could in
principle be associated with the lens was also detected, the light
curve parameters imply that the lens must be only $\sim 300\,$pc from
the source if it is responsible for this excess flux.  The {\it a
priori} probability of such a small lens source distance is comparable
to the {\it a priori} probability that the excess flux is due to a
binary companion to the source or lens star, or even a completely
unrelated star. So, this event is ambiguous \citep{moa310}

The situation is quite different for OGLE-2006-BLG-109. In this case,
it is very unlikely that the excess flux superimposed on the source is due
to a star other than the lens/planetary host star for a couple of different
reasons. First, the excess flux in $H$ matches the prediction for
the mass and distance of lens star as determined from the microlens
parallax and finite source effects, and second, it is 
brighter than bulge main sequence stars, the number of stars the
observed brightness is relatively low. Nevertheless, it
is a logical possibility that the planet host could be a dim white dwarf
and the detected flux could be from another star, although we note 
that despite several systematic searches \citep{wd_pl1,wd_pl2,wd_pl3},
the only planet known to orbit a white dwarf was apparently 
involved in a complicated dynamical interaction in the core
of a globular cluster \citep{pulsar_wd_pl}. 
We now proceed to estimate the probability that another star
besides the planetary host star could be responsible for the detected
$H$-band flux using arguments similar to those of 
\citet{moa310}. There are three possibilities: the chance superposition
of an unrelated star or a binary companion to the
lens or source star.

The most likely alternative is the chance superposition of an unrelated star.
The density of stars
within 0.4 mag of the $H$-band flux attributed to the lens star
is 0.126 stars per square arcsecond, so the probability that 
such a star is close enough to the source so that the source is not
separately detected in the best seeing $K$-band image is 0.3\% assuming
that the star can just be detected at a separation of 1-FWHM.

The next possibility is that the excess $H$-band flux could be from a binary
companion to the source star. Unlike the case of MOA-2008-BLG-310
\citep{moa310}, the companion would have to be substantially brighter
than the source. It would be 1.7 mag brighter than the 
source in the $H$ band, which would make it brighter than the 
top of the bulge main sequence \citep{holtz,zoccali}. With the same
extinction and distance as the source, the absolute $H$-band magnitude
of this companion would be $\sim 2.05$, which would imply that it
should be a hydrogen-shell burning star just beginning its rise up the 
red-giant branch. Assuming a typical bulge age of $\sim 10\,$Gyr, it
would have a mass of just over a solar mass and about
10\%-15\% more massive than the source. 
Figure 10 of \citet{bin_gdwarf} implies that about 1.4\%
of G-dwarfs should have a binary companion that is 1-1.1 times more
massive. However, this phase of stellar evolution is quite short-lived. 
From \citet{sun_evolve}, we estimate that the Sun will have an 
absolute $H$ magnitude of $2.05 \pm 0.3$ for a time period equal to
only about 2\% of its main sequence lifetime.
Furthermore, a star's lifetime varies roughly as the $-2.5$ power of
the mass, so the initial mass interval corresponding to bulge stars
with an absolute $H$-band magnitude of $2.05 \pm 0.3$ is
only $0.08\msun$. This means that
only 8\% of the bulge G-dwarfs with companions
in the 1-$1.1\,\msun$ mass range should be in this magnitude range. Finally,
based on Figure 7 of \citet{bin_gdwarf}, we expect that 25\% of these
binary companions would have a separation of $> 700\,$AU where they
could be detected in the best Keck images, and another $17\,$\% would
have an orbital period of $< 200\,$days which would imply a detectable
``xallarap" (source orbital motion) effect in the microlensing light curve.
So, the {\it a priori} probability of a source companion with the observed
$H$-band magnitude is only 0.06\%.

The final alternative is a companion to the lens. This possibility is 
excluded by the lack of any signature of an additional
companion in the light curve. In this scenario, the companion
must be at the same distance as the lens, so the $H$-band 
brightness means that lens must be a white dwarf of $\sim 0.5\,\msun$,
and the companion must be an M-dwarf of essentially the same
mass. The light curve shows no evidence of such a companion, so
we use the method of \citet{mps-98blg35} to work out the
constraints on an additional lens companion at the same mass as
the lens star, and this analysis indicates that such companions
separated by less than $0.18^{\prime\prime}$ from the 
primary lens star are excluded. However, the lens and source 
should be separately detected in the best $K$-band image if
the separation is $> 0.09^{\prime\prime}$, so this possibility is
ruled out.

So, the {\it a priori} probability that source of the excess flux
observed in the Keck AO images is something other than the lens
star is less than 0.4\% under the assumption that white dwarfs are
as likely to host planets as early M-dwarfs.

\subsection{Lens Brightness Constraint}
\label{sec-lens_bcont}

Now that we have shown that it is extremely unlikely for the
excess $H$-band flux superimposed on the source star to belong to another
star, beside the lens star, we will make the assumption that the
flux we detect is from the lens star. Our measurement gives
$H_L = 17.17 \pm 0.05$, but at its likely distance of
$D_L \approx 1.5\,$kpc, the lens is only located about $35\,$pc from
the Galactic plane, and so
the extinction toward the lens is likely to be substantially
less than the extinction toward the source. This means that
we cannot use all the Galactic bulge stars within a small
angular distance from the lens star to estimate the 
foreground extinction.

The distribution of the dust in the Galactic disk is somewhat complicated
\citep{drimmel,marshall_blg_ext06}, and the distribution along the 
line of sight at a distance of $D_L \approx 1.5\,$kpc toward the
Galactic center is not very well known. So, we will use a very
simple model, with generous error bars. At $1.0\,$kpc interior to
the solar circle, the scale height of the dust distribution is about 
$110\,$pc. At a Galactic latitude of $b = -1.8901^\circ$, the decrease
in dust density due to the line of sight leaving the disk plane is
essentially canceled by the increase in density due to the 
approach to the Galactic center. However, some models 
have a flared disk
that decreases the scale height at smaller Galactocentric radii. Also,
the dust distribution is not thought to continue all the way into the 
Galactic bulge. To get a crude estimate of the extinction, we simply
assume a scale height of $110\,$pc that continues all the way to
the source. For source extinction of $A_{HS} = 0.60$, this gives
$A_{HL} = 0.24 \pm 0.24$, where we have assumed a large
uncertainty because of the crudeness of this estimate. Of course, we
do not allow $A_{HL} < 0$, and $A_{HL} = 0$ should be quite unlikely.
But, at these coordinates, some of the dust models imply that the
dust density increases all the way to the bulge. So, it is quite
possible that only a very small fraction of the extinction is in the
foreground of the lens star.
Combining this
with the estimated lens star $H$-band brightness, we find a 
dereddened lens star brightness of 
\begin{equation}
H_{L0} = 16.93 \pm 0.25 \ ,
\label{eq-Hl0}
\end{equation}
which we can use to compare to the predictions of the lensing models.

\subsection{Magellan Spectrum and Host Star Kinematics}
\label{sec-spec}

A spectrum of the lens star, the source star, and the 
bright star $0.35^{\prime\prime}$ to the North of the lens and source
was obtained using the MIKE Spectrum on the
Magellan Telescope by G.~Pietrzy{\' n}ski on 2007 March 25. 
This spectrum covered the spectral range 5400-$6500$\AA.
It indicates
that this bright neighboring star is indeed a red clump giant
star as its positions in the CMDs 
imply. (See Figures~\ref{fig-cmd_VI}-\ref{fig-cmd_IH}.)
A cross-correlation analysis of this spectrum was kindly provided
by Ian Thompson and Andy McWilliam of the Carnegie Observatories,
using HD193901 (a metal poor subgiant) as the template. This
cross-correlation analysis was done with the IRAF FXCOR package,
and used the spectral ranges 5400-5875\AA\ and  5925-6500\AA\ to 
avoid strong interstellar absorption in the Na-D lines. The resulting
cross-correlation function clearly shows two peaks. The strongest
peak has a Heliocentric radial velocity of $v_r = 125\,$km/s, 
and the second strong peak has a Heliocentric radial velocity of
$v_r = -49\,$km/s. The peak ratio was about 2.6, and the formal
uncertainties reported by the IRAF FXCOR package was less than
$1\,$km/s for each peak. We attribute the strongest peak to the
bright blend star, and the second strongest peak to the planetary
host star. The source star is too faint to be detected by this
analysis.

The radial velocity of the blend star is consistent with our classification
of it as a bulge giant, but it is not otherwise interesting. The kinematics of
the host star can be compared to those of some 528 stars 
with $M_V > 4$ within $25\,$pc
of the Sun as compiled by \citet{reid_mdwarf_kin}. Converted from the
Heliocentric frame to one at rest with the average of this nearby star
population, we find that the radial velocity of the host star is 
$v_r = -36\,$km/s, which has an absolute value slightly smaller
than the RMS value of $40\,$km/s.

We do not have a direct measure of the transverse motion of the host
star, but the microlensing parallax and finite source
measurement does give us the relative velocity between the host
and source stars. Since the distance to the host, 
$D_L \approx 1.5\,$kpc is much smaller than the distance to the 
source, $D_S$, this tells us mostly about the host star.
If we assume that the source is a bulge star, and the bulge has no
rotation and has a velocity dispersion of $80\,$km/s in both the 
Galactic North and rotation directions, then we find that the
velocity distribution that we infer for the host star is 
$(v_V,v_W) = (-45,-19) \pm (15,15)\,$km/s, where $v_W$ and $v_V$
are the components of the lens velocity in the Galactic north and
rotation directions. Here, the uncertainty is entirely due to the 
bulge velocity dispersion. The velocity dispersion for the local
star sample \citep{reid_mdwarf_kin} is 
$(\sigma_W,\sigma_V) = (28,19)\,$km/s. We can add this in quadrature
to the uncertainty due to the unknown source star velocity, and we
find $(v_V,v_W) = (-45,-19) \pm (32,24)\,$km/s. So, two of three components
of the host star velocity are within 1-$\sigma$ of the expected value and the
third component is 1.4-$\sigma$ behind the mean velocity in the
rotation direction. So, the kinematics are quite consistent with a bulge source
and a relatively old planetary host star in the disk, as we would expect.

\section{Alternative Lens Models}
\label{sec-alt}

High magnification planetary microlensing events are
known to have a number of approximate parameter degeneracies
that can often complicate the interpretations of these 
events. We explore these alternative models with respect to the
reference model with $\chi^2 = 2542.06$ for 2557 degrees of 
freedom. (See Table~\ref{tab-pparam} for the parameters of this model.)
The first of these degeneracies is the well known
$d \leftrightarrow 1/d$ degeneracy \citep{griest_saf,dominik99} that applies to 
high magnification events unless $d \approx 1$. In
the case of OGLE-2006-BLG-109, this would apply to the 
Jupiter-mass planet at $d \approx 0.6$ (or $d\approx 1.6$), but not
to the Saturn-mass planet at $d = 1.04$. This Saturn-mass planet
is a case of a so-called ``resonant caustic\rlap," in which the planetary
caustic is connected with the central caustic. Since the source
trajectory encounters the planetary part of the caustic, the
$d \leftrightarrow 1/d$ degeneracy is broken for this planet.
In \citet{gaudi-ogle109}, we reported that the degenerate
solution with $d_J \approx 1.6$ was disfavored by $\Delta\chi^2 = 11.4$,
which was enough for us to formally exclude this solution.
However, the situation is a bit more complicated in the current
analysis, as we describe below.

The other approximate degeneracies concern the microlensing
parallax effect. These have been discussed in the context of
single lens events by \citet{smith_par_acc}, \citet{gould-jerkpar}, and
\citet{multi-par}. The symmetries of the first few
terms of the Taylor expansion of the parallax effect )with respect to
time) lead to approximate degeneracies of the full effect for most
events, which have $t_E \ll 1\,$yr. However,
one of the symmetries of single and double-lens models is removed
by the third lens. High magnification planetary microlensing events often
have four degenerate parallax solutions \citep{bennett-moa192}, but an 
additional planetary mass ruins the reflection symmetry about the
lens axis (because there is no longer a unique lens axis). As a 
result, there is only a single discrete degeneracy that remains.

With the modeling runs used for the OGLE-2006-BLG-109 discovery
paper, there was a degenerate model with parameters similar to
those listed in Table~\ref{tab-mparam} except that $u_0 = -0.00344$,
$\theta_{1\rm cm} = -2.5266$, $\phi_E = 0.684$, and $\pi_E = 0.239$. 
This model was disfavored compared to the best fit model presented
in that paper by $\Delta\chi^2 = 7.6$. Although this $\Delta \chi^2$ alone is formally
enough to exclude this model, it was further disfavored because the
change in $\pi_E$ value implies a much larger lens system mass of $M_L
\approx 0.77\msun$ at a somewhat larger distance of $D_L = 2.06\,$kpc,
which would predict an $H$-band brightness
that is $\sim 0.8$ mag brighter than the observed $H$-band 
magnitude, which corroborates the rejection of this model.

The rejection of this alternate model is further confirmed by our inclusion
of the terrestrial parallax effect in our current modeling.
This is simply the parallax effect due to the locations of the different
observatories on the surface of the Earth.  In our
previous modeling, we effectively assumed that all observations take place
from the center of the Earth. With the inclusion of this effect, the best fit
$\chi^2$ improves by $\Delta\chi^2 = 13.5$, but the $\chi^2$ difference 
with the $u_0 < 0$, $\theta_{1\rm cm} \approx -2.5$, $\phi_E \sim 0.7$
model increases to $\Delta\chi^2 = 39.3$. So, we can consider this 
model to be strongly excluded by several lines of evidence.

However, the situation is more complicated with the $d_J > 1$ model. When
terrestrial parallax is added to this model, the
$\chi^2$ of this model improves by more than the $d_J < 1$ model, so that
the $\chi^2$ difference between these models drops to $\Delta\chi^2 = 1.8$.
Thus, the model with the Jupiter orbiting outside the Saturn is no longer formally
excluded by this $\chi^2$ difference. However, as we shall discuss in
\S~\ref{sec-orbits}, this best fit $d_J > 1$ model is almost completely
inconsistent with any co-planar stable orbit, so the conclusion of 
\citet{gaudi-ogle109} that the Jupiter orbits inside the Saturn is still 
likely to be correct.

\section{Converting Model Parameters to Physical Orbits}
\label{sec-orbits}

As we mentioned in \S~\ref{sec-constrain}, there are two possible 
interpretations of our orbital motion modeling scheme. The first
interpretation is to take the circular orbit model literally. This is
an effective way to ensure that the orbital model parameters do
not stray into unphysical regions of parameter space. But, of course,
it is unlikely that the orbit of the Saturn-mass planet is precisely
circular, but our orbital model allows only a circular orbit. 

In order to extract the set of physical orbits that are consistent with
the parameters of a given model, we adopt the second interpretation
of our orbital parameter models. In this interpretation, the 
$1/T_{\rm orb}$ parameter is considered to be a measure of the
orbital acceleration, to which it is equivalent at second order.
With this interpretation, we compute the acceleration implied by
the circular orbit model. Then, using the measured $\theta_*$ value 
and eqs.~\ref{eq-mass} and \ref{eq-Dl} we can determine the relative
distance along the line of sight between the Saturn-mass planet and 
its host star. With the other fit parameters, this gives us the 
three-dimensional position and the velocity in the plane of the sky.
Since the mass is known from eq.~\ref{eq-mass}, this leaves
one additional parameter, the velocity along the line of sight, to
describe the orbit. Thus, we have a single parameter 
family of orbits for each set of model parameters. 
Given a set of fit parameters, we first calculate
the minimum energy of the orbit using the three-dimensional 
position of the planet and the two dimensional velocity. If
this energy is positive, then there are no bound orbits 
consistent with these fit parameters. If the energy is negative,
then we can construct the family of orbital solutions by varying 
the velocity along the line of sight from the maximum value
consistent with a bound orbit to the negative of that value.

It may seem odd that we can use a model of describing a circular
orbit can be interpreted to describe a one-parameter family
of general orbits, because it normally requires two more parameters
to describe an orbit that is not constrained to be circular. However,
if we assume that the distance to the source is known, the 
circular orbit problem is over-constrained by model parameters,
the application of Kepler's third law, and the measurement of the
source star angular radius, $\theta_*$. We avoid this 
problem of too many constraints, by considering $D_S$ to be
a variable, with the soft constraint, eq.~\ref{eq-chi2Dl}. Now,
when we generalize to non-circular orbits, we fix $D_S$,
and use the additional constraint to avoid an additional non-circular
orbit parameter.

The duration of the light curve signal of the Jupiter-mass planet is
too brief to reveal any information regarding its orbit. However,
it is well accepted that planet formation occurs in a disk, so that
all planets are expected to form in one plane. Furthermore, the only
known exoplanet systems that have measured inclinations of
more than one planet are the solar system and the two
most massive planets of the PSR 1257+12 system
\citep{pulsar_planets03}. In both cases the planets orbit
in nearly the same plane. Thus, it is natural to assume that
the planets in the OGLE-2006-BLG-109L system are also
co-planar. If so, we can use this assumption to provide tighter
constraints on the orbital parameters of the Saturn-mass planet,
OGLE-2006-BLG-109Lc.

In addition to requiring that the Saturn-mass planet has parameters
that correspond to a bound orbit, we can also require that its orbit
and that of the Jupiter-mass planet are in a stable configuration. 
With the assumption that the orbits are co-planar, we can define a 
unique three-dimensional position for the Jupiter, once the orbit of
the Saturn-mass planet has been defined. As described above, this
requires that we specify a line-of-sight velocity for the Saturn-mass planet
in addition to the parameters that describe the light curve model.
With only the position of the Jupiter-mass planet specified, we have
a lot of freedom for the rest of the orbital parameters. However, the
choice that will impose the weakest orbital stability constraints on
the orbit of the Saturn-mass planet is simply to assume a circular 
orbit for the Jupiter-mass planet, so this is what we assume.

With the orbits of both planets now specified, we can now check for
orbital stability. The simplest constraint is to insist that the
orbits do not cross, but orbits do not have to cross to be unstable.
We employ the analytic Hill stability criterion of \citet{barnes_stab06} as
our condition for orbital stability. While the majority of unstable orbits
are removed by the simple orbit crossing criterion, it is
only the Hill stability criterion that can detect unstable circular 
orbits, and it is in fact this latter condition that excludes the best
fit $d_J > 1$ model discussed in \S~\ref{sec-alt}. In particular, only
0.04\% of the parameter sets in our MCMC chain calculations which
corresponded to stable orbits with $d_J > 1$. However, some of the
$d_J < 1$ Markov chains were much longer than the $d_J > 1$ chains,
and other $d_J < 1$ chains had $T_{\rm orb}$ values with relatively
large $\chi^2$ values, while their $d_J < 1$ counterparts generated
no systems with stable orbits. As a result, these stable orbit, $d_J > 1$, 
receive a higher than average weighting, but they still account for
only 0.3\% of the total weight. Thus, the $d_J > 1$ models are excluded
at 99.7\% confidence.

\section{Bayesian Analysis of the OGLE-2006-BLG-109L Planetary System}
\label{sec-param}

In \S~\ref{sec-orbits}, we described how we could convert a single 
set of light curve model parameters into a one-parameter family of
complete orbital solutions for both planets. We now apply this method
to determine the orbital parameters of the OGLE-2006-BLG-109Lb,c
planetary system. We do not apply the circular orbit constraint for the
Saturn-mass planet to these
calculations, but we do constrain the brightness of the lens star using
the mass and distance derived from eqs.~\ref{eq-mass} and \ref{eq-Dl},
and dereddened $H$-band brightness of the lens star given by
eq.~\ref{eq-Hl0} using an analytic model of the $H$-band 
mass-luminosity relation based on 
\citet{kroupa_tout}, 
\citet{delfosse} and 
\citet{henry_mc}.
Our basic method here is the MCMC, but there
are several complications that must be addressed. 

The first complication is that there are several relatively flat directions
in parameter space that are not efficiently probed by the MCMC calculations,
and efficiency is quite important because these calculations are rather
CPU intensive. We perform a set of MCMC calculations with the 
fifth orbital parameter, $T_{\rm orb}$ (the orbital period of
the assumed circular orbit) fixed to a series of values:
2000, 2500, 2857, 3333, 4000, 5000, 6667, 10000, and 20000 days.
These runs were done for both $d_J < 1$ and $d_J > 1$ models, but 
most of the $d_J > 1$ produced no stable orbits. The only exception
was the $d_J > 1$, $T_{\rm orb} = 10^4\,$days run, which 
yielded a few stable orbital solutions, although these amounted to
only about 0.03\% of the orbits that were consistent with the fit
parameters of this MCMC run. In contrast, the run with $d_J < 1$ and
$T_{\rm orb} =5000\,$days, which is closest to our best-fit
circular orbit model had a 50\% stability rate for orbits that were
consistent with the model parameters or the MCMC run.
Also, the runs with $T_{\rm orb} $ at the extremes of the range 
considered had few acceptable orbits. The run with
$d_J < 1$ and $T_{\rm orb} =20,000\,$days produced no
bound Saturn-mass planets, and the $d_J < 1$ with 
$T_{\rm orb} =2000\,$days
produced only 0.6\% bound orbits. The best fit model at this $T_{\rm orb} $
value is also disfavored by $\Delta\chi^2 = 7.5$ compared to the
best fit $T_{\rm orb} =2000\,$day model, and $\chi^2$ grows rapidly
with smaller $T_{\rm orb}$. So, our range of 
$2000\,{\rm days} \leq T_{\rm orb} \leq 20,000\,{\rm days}$ seems to cover the
range of viable orbital accelerations. These different MCMC runs
with different fixed $T_{\rm orb}$ values are combined with different weightings
of $e^{-\Delta\chi^2/2}$ based on the best fit $\chi^2$ for each $T_{\rm orb}$ value.

There are additional uncertainties that are not accounted for in these
MCMC runs, so we have included them in our integration over the 
MCMC results. As mentioned in \S~\ref{sec-orbits}, the model parameters
only specify five of the six parameters needed to completely
describe the orbits, so we
integrate over all the orbits consistent with the orbital stability constraint 
for each link in each Markov chain. Also, the uncertainty in the derived
angular radius of the source star, eq.~\ref{eq-thetaS}, is not included in
the MCMC runs, so we select $\theta_*$ from a Gaussian random
distribution based on eq.~\ref{eq-thetaS} when we sum the
results of the MCMC runs.

The MCMC runs also do completely sample the microlensing 
parallax parameter space due to the partial degeneracy between
the orbital motion and parallax parameters. In order to ensure that 
the uncertainty in $\pi_E$ is not underestimated, we have found the
best fit models with $\pi_E$ fixed on a one-dimensional grid with no
lens brightness constraint. This calculation yields results that are
well fit by $\pi_E = 0.338\pm 0.037$. We add this as an additional
uncertainty in the same way as the uncertainty in $\theta_*$.

One consideration that has very important consequences for the
orbital parameter results is the proper application of the Bayesian
prior distribution for the orbital parameters. The model parameters
have been designed for convenience and calculational efficiency
during the modeling calculations, but they also implicitly provide
the Bayesian prior for the MCMC calculation, which is basically 
a Bayesian likelihood calculation. In fact, the Bayesian prior provided by
these parameters is quite an unreasonable one. This must be
corrected to a more reasonable prior distribution by computing the
Jacobian determinant of the coordinate transformation between
the parameters used for the MCMC calculation and a set of parameters
that describes a reasonable prior distribution. We select a prior
that is uniform in orbital phase, eccentricity, time of periapsis, and 
the logarithm of the semimajor axis and has a random orientation
of the orbit. The derivatives needed for the Jacobian determinant 
are calculated numerically, and the results have been shown not to
depend on the numerical derivative calculations. However, to 
avoid numerical difficulties with this calculation, we have had 
to exclude orbits with orbital semimajor axes $> 800\,$AU.
Virtually all the models in the MCMC runs must have 
three-dimensional star-planet separations
that are $< 10\,$AU, and there are no viable 
models with three-dimensional separations
much larger than $10\,$AU, because the light curve requires planet
velocities that would result in an unbound planet if the 
three-dimensional separation
at the time of the event was $\gg 10\,$AU. So, these excluded models
with $a > 800\,$AU all have the planet detected extremely close to 
periapsis. which makes them extremely unlikely, and thus their
exclusion has no effect on the results (except to remove large 
numerical errors).

\subsection{Physical Parameters of the OGLE-2006-BLG-109L System}
\label{sec-params}

The results of our MCMC calculations and
Bayesian planetary parameters analysis are
summarized in Tables~\ref{tab-mparam} and \ref{tab-pparam}.
Tables~\ref{tab-mparam} shows the nonlinear
fit parameters of the
best fit circular orbit model and the mean and RMS of the 
models contributing to the MCMC Bayesian analysis. The subscripts
1, 2, and 3 refer to the Jupiter-mass planet, the Saturn-mass planet,
respectively.
Table~\ref{tab-pparam} shows the median, and the 1-$\sigma$
and 2-$\sigma$ limits on the physical parameters of the planets
and their host star. $D_L$ is the distance to the lens system.
$M_A$, $m_b$, and $m_c$ are the masses of the host star,
and the Jupiter and Saturn, respectively. The semimajor axes
and periods are given by $a_b$, $a_c$, $P_b$, and $P_c$.
For the Saturn-mass planet, $c$, we also constrain the
orbital eccentricity, $\epsilon_c$, the inclination angle,
$i_c$, and the axis of inclination, $\alpha_c$. The
final two parameters, $K_b$, and $K_c$, 
are the velocity amplitudes that are
normally measured by the Doppler radial velocity method.

This analysis shows that the Saturn-mass planet, OGLE-2006-BLG-109Lc,
is most consistent with being in a low-eccentricity orbit, like Saturn, in our own solar
system, but the constraints are not very tight, with 1-$\sigma$ and
2-$\sigma$ upper limits on $\epsilon_c$ of 0.32 and 0.62 respectively.

Note that the parameters in this table assume that the Jupiter-like
planet is in a circular orbit. The uncertainty in $a_J$, the semimajor
axis of the Jupiter would undoubtably be larger without this assumption.
However, the constraints on the eccentricity, $\epsilon_c$, 
of the Saturn-like planet would probably be tighter and would likely favor
somewhat smaller eccentricities, if we had allowed non-circular orbits
for the Jupiter. The reason for this is that an eccentric orbit for the Jupiter
would make stability more difficult for the Saturn, and would likely exclude
some of the higher $\epsilon_c$ orbits that are (barely) allowed in the
current analysis.

We have also done a similar analysis of the properties of the Saturn-mass
planet without the assumption that the Jupiter-mass planet is in
a coplanar orbit, and the results are presented in Table~\ref{tab-pparam_noco}.
These results are similar to, but slightly weaker than,
the results with the coplanar orbit assumption shown in Table~\ref{tab-pparam}.
This indicates that most of the constraints on the Saturn-mass planet's
orbit are likely to be due to the velocity components in the plane of the
sky that have been measured with some precision from the 
microlensing light curve.

Some characteristics of the Saturn-mass planetary orbits that are 
consistent with the light curve are shown in Figure~\ref{fig-eps_gam},
which shows the probability distribution in the eccentricity,
$\epsilon_c$, versus inclination, $i_c$, plane. Some of the
structure in this plot is a result of the computational shortcut we have
taken by running our MCMC runs with the model parameter
$T_{\rm orb}$ fixed at a fairly sparse set of values. However, the
highest probability peaks in this distribution at $\epsilon_c \simeq 0$
and $i_c \simeq 56^\circ$ and  $i_c \simeq 66^\circ$
have a different cause. These represent the two orbital configurations
for which the most likely two-dimensional velocity of the
Saturn-mass planet can match a circular orbit \citep{dong_ogle71}.

\section{Limits on Additional Planets and Lens Stars}
\label{sec-limits}

We have determined limits on other planets in the OGLE-2006-BLG-109L
system using the method of \citet{mps-98blg35}. In order to properly
apply this method, we must first check that the fit $\chi^2$ cannot
be significantly improved by adding an additional planet. 
At present, there is no existing modeling code that can handle a
lens with 4 masses and orbital motion, so do we not attempt such
fits. Instead, we have inspected the peak region of the light curve
and determined that there is no localized region of the light curve
where an additional, localized planetary signal could significantly
improve the fit. Based on this, we have applied the 
\citet{mps-98blg35} method with a threshold of $\Delta\chi^2 \geq 60$.
This allows us to exclude a
Neptune-mass ($17\mearth$) planet in an annulus of
projected separations ranging from 1.1-$4.5\,$AU. If we assume
coplanar orbits with the median $i_c = 64^\circ$ inclination, 
we can de-project this to give an elliptical
annulus that extends out to 2.5-$10.3\,$AU. However, most of this
exclusion region is probably already excluded by orbital stability
constraints. This will generally be the situation for low mass planets in
systems with gas-giant planets discovered near the Einstein radius.

The exclusion regions for additional giant planets are substantially larger.
We can rule out another ``Jupiter\rlap," with the same mass as
OGLE-2006-BLG-109Lb over the projected separation range
0.5-$10.5\,$AU. When we de-project this to get the exclusion region
in the plane of the orbit, we find that the exclusion region extends
from $1.1\,$AU to $24\,$AU at the widest part of the elliptical 
annulus. The exclusion region rules out another ``Jupiter" orbiting
at a bit more than twice the orbital distance of the Saturn-mass
planet. The exclusion region for another ``Saturn\rlap," with the same mass as
OGLE-2006-BLG-109Lc, extends about 72\% as far as the exclusion region
for another ``Jupiter\rlap".

These limits on additional planets with a mass greater than 75\% of
Saturn's mass are essentially in the constant
shear limit \citep{chang-refsdal89}, where the light curve deviation
depends primarily on the shear parameter, $\gamma = q/d^2$,
and for this event are limit corresponds to $\gamma < 7\times 10^{-5}$.
This is the reason for our exclusion (in \S~\ref{sec-excess})
of the possibility that the
$H$-band flux attributed to the primary lens star could be from
a companion to the lens, instead.

\section{Discussion and Conclusions}
\label{sec-conclude}

We have presented the complete analysis of the OGLE-2006-BLG-109Lb,c
planetary system that was summarized by \citet{gaudi-ogle109}, and we have
explored the physics implications of the microlensing fit parameters.
We have introduced a new method to determine the angular radius of the
source star, $\theta_*$, which makes use of three color, $VIH$, light curve
data to determine $\theta_*$ to a precision of 2.8\%. We then combine this
measurement with the microlensing parallax parameter, $\pi_E$, determined from
the microlensing light curve model to yield a direct, geometric, measurement of
the masses of the star and two planets in this planetary system. This is
the only multiple planet system with measured masses for the
stars and planets.  (The pulsar planet system PSR B1257+12
\citep{pulsar_planets03} has measured planet masses, but the mass
of the host neutron star is assumed.) The mass
measurement of the host star is confirmed by Keck AO images, which detect
the planetary host star, and measure its $H$-band magnitude to be
$H_L = 17.17\pm 0.05$.

These results, plus our more sophisticated modeling, including terrestrial
parallax, confirm the results of \citet{gaudi-ogle109} that this system
resembles a smaller version of our own solar system with 
a primary about half the mass of the Sun orbited by planets slightly
smaller than Jupiter and Saturn in a similar arrangement with
the Saturn-like planet outside. Their semimajor axes are about half of
those of Jupiter and Saturn.
This similarity to Jupiter and Saturn is
even greater if we consider the configuration of the system during the
process of planet formation
at an age of $\sim 1\,$million years, when nuclear fusion does not 
yet dominate the host star's energy production. At this time,
the stellar luminosity is thought to
scale as $\sim M^2$
\citep{burrows93,burrows97} (G. Kennedy, C. Lada, 2007, private communications).
In the core accretion theory, the most massive planets are thought to form
outside the ``snow line" where it is cold enough for water-ice to form
\citep{ida_lin,lecar_snowline,kennedy-searth,kennedy_snowline}, and
this predicted $\sim M^2$ dependence of the proto-star luminosity 
implies that the location of the ``snow line" should scale linearly with
stellar mass. So, if our solar system's snow line was at $2.7\,$AU,
which is slightly larger than half of Jupiter's semimajor axis, then we
estimate that the snow line for OGLE-2006-BLG-109L should have been at
$\sim 1.4\,$AU, which is slightly larger than half the semimajor axis of its
Jupiter-analog planet.

Due to the relatively long, 11-day, duration of the signal of the Saturn-analog
planet, the light curve tightly constrains four of the six parameters that are
needed to describe the planetary orbit. A fifth orbital parameter is weakly
constrained. We use these light curve constraints in an MCMC 
analysis, along with orbital stability constraints for the
combined Jupiter/Saturn-analog system,
to determine the constraints on the orbital parameters implied
by the light curve model. Assuming coplanar orbits, we find an orbital
inclination of $i_c = 64^\circ {+4^\circ\atop -8^\circ}$, with
a 2-$\sigma$ lower limit of $49^\circ$. Because the lens star is relatively
bright and brighter than the source, 
this means that it should be possible to confirm at least the
inner, Jupiter-analog, planet with Doppler radial velocity measurements.
While, the host star, at $H_L = 17.17\pm 0.05$ is much fainter than the host
star for any known planet discovered with the radial velocity method,
it seems reasonable to expect that the next generation of very large
telescopes \citep{tmt,e-elt-sci,e-elt-stat}, combined with a 
high-throughput, high resolution spectrograph \citep{mayor_harps}
would be able to detect the radial velocity signal of the
Jupiter-analog planet with a radial velocity amplitude of
$K_b = 17.4 {+1.3\atop -1.1}\,$m/sec. This would be a challenging measurement
for these extremely large telescopes, but our prior knowledge of the
planetary system will minimize the number of observations that would be
required. It is likely to be at least a decade and perhaps two decades
before such measurements are possible, but this is much sooner than
the $\simgt 10^6\,$yr that we would likely have to wait before either of
these planets reveals themselves again via microlensing.

With our measurements of the planetary masses and constraints on
the orbital parameters, OGLE-2006-BLG-109Lb,c is arguably the 
best constrained multiplanet system to orbit a main sequence star,
other than the Sun. None of the other known multiple exoplanet systems has
measured masses, and radial velocity measurements generally do
not constrain more parameters of the planetary orbits than we have
constrained for OGLE-2006-BLG-109c. On the other hand, radial
velocity surveys often measure the eccentricity of the planets they
detect more precisely
than we have done, and the eccentricity is intrinsically more interesting
than the parameters we have measured describing the orientation of
the orbital plane of a single planet. Nevertheless, this is certainly
the first planetary system found by microlensing in which the eccentricity
of a planetary orbit is constrained.

\begin{figure}
\plotone{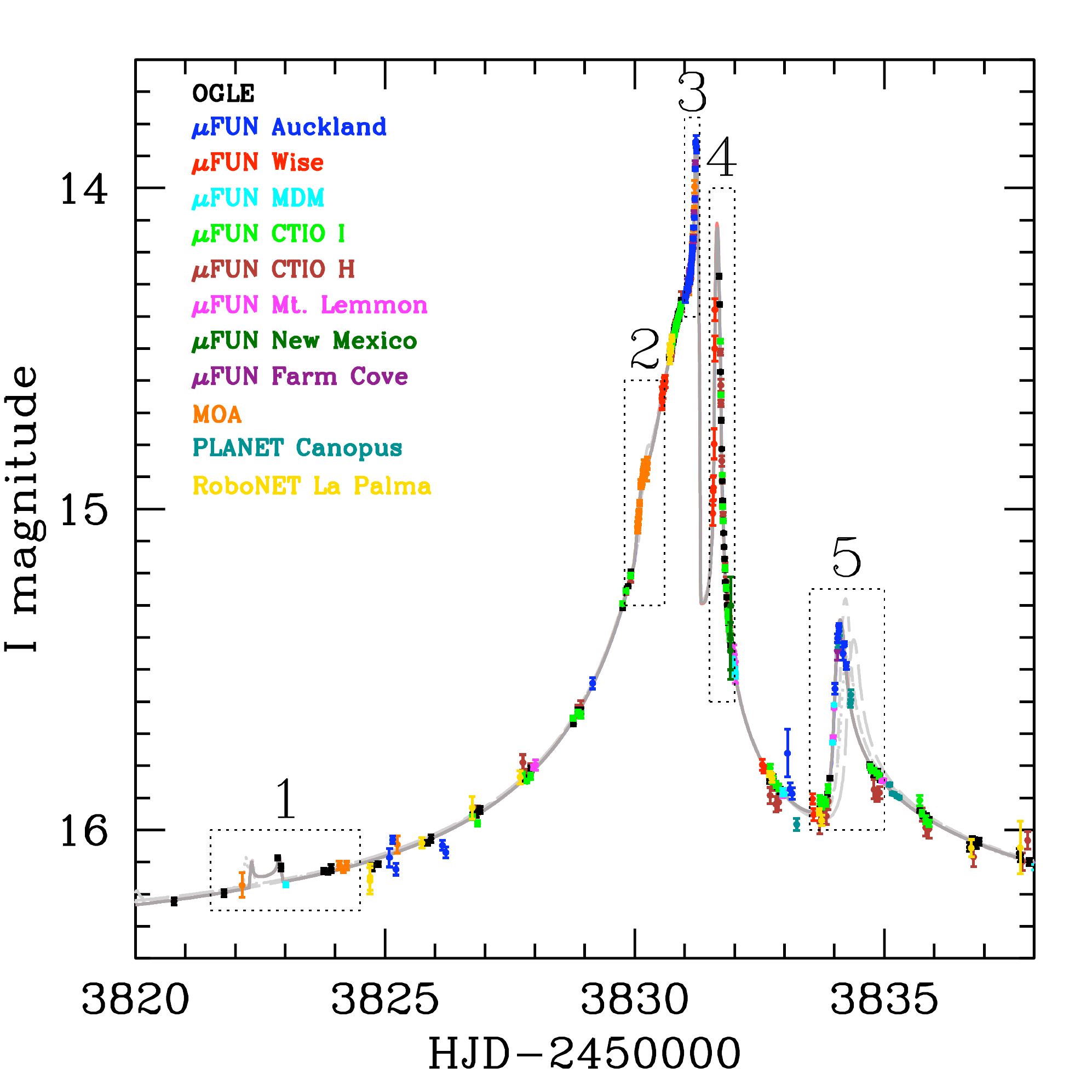}
\caption{Photometric measurements and model light curves are
shown for 18 days near the peak of the
OGLE-2006-BLG-109 light curve, with telescope and
passband for each data point indicated by its color, as indicated.
The best fit model is plotted as a solid gray curve, and the light
gray curves plotted with dots, short-dashes, and long-dashes 
indicate the same model, but with parallax, orbital motion,
and both parallax and orbital model removed, respectively.
(These alternative models are more easily seen in Figures
\ref{fig-lc1} and \ref{fig-lc2}.)
The five caustic crossing and cusp approach or crossing
features are indicted by the dotted boxes labeled 1-5.
\label{fig-lc1}}
\end{figure}

\begin{figure}
\plotone{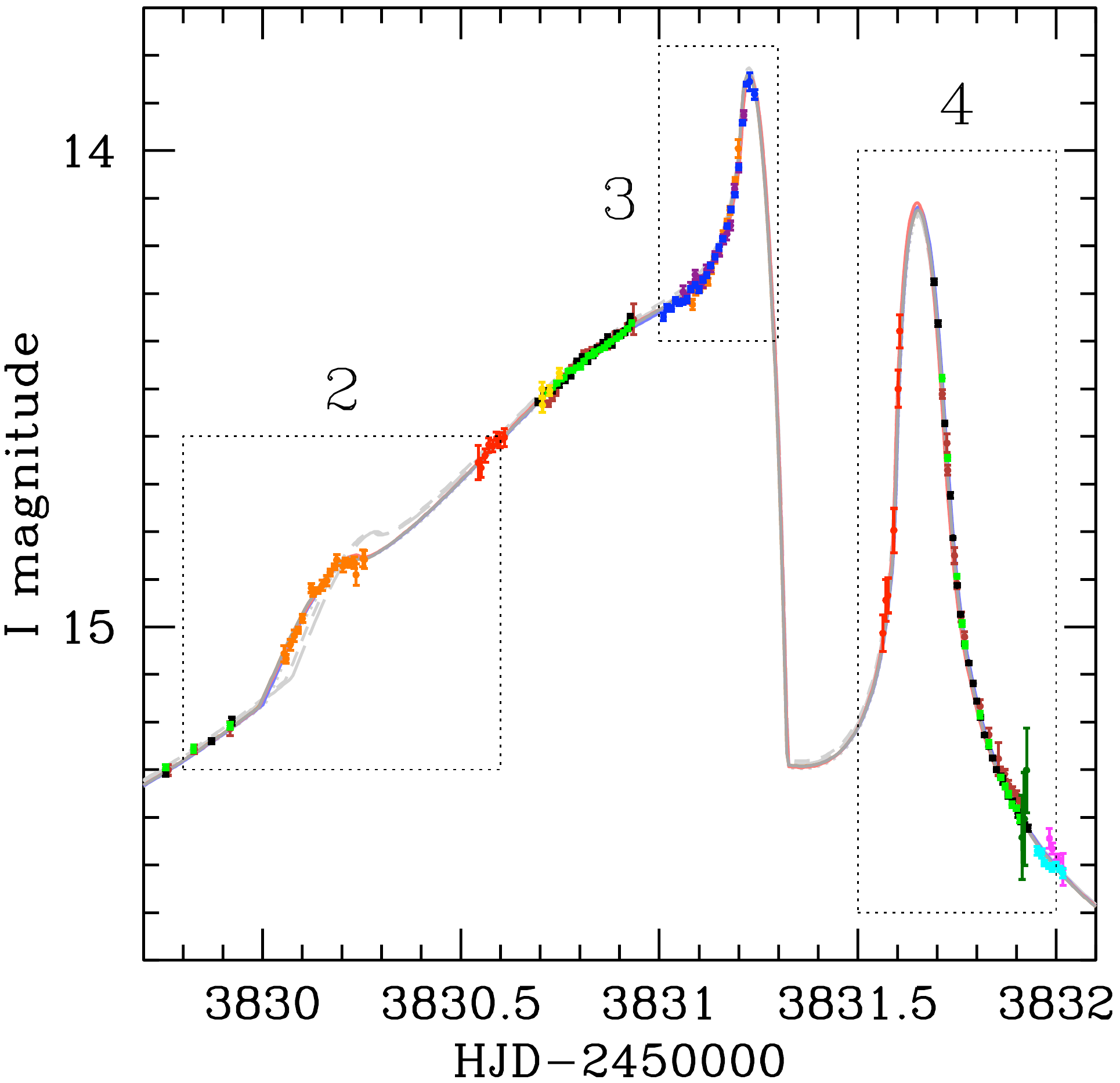}
\caption{Close-up of the OGLE-2006-BLG-109 light curve peak 
showing caustic crossing and cusp approach features 2-4.
OGLE-2006-BLG-109 light curve peak. As in Figure~\ref{fig-lc1}, the
light gray curves plotted with dots, short-dashes, and long-dashes 
indicate the best-fit model without parallax, orbital motion, and both
parallax and orbital motion, respectively, while the solid curve(s)
indicate the best fit model.  
\label{fig-lc2}}
\end{figure}

\begin{figure}
\plotone{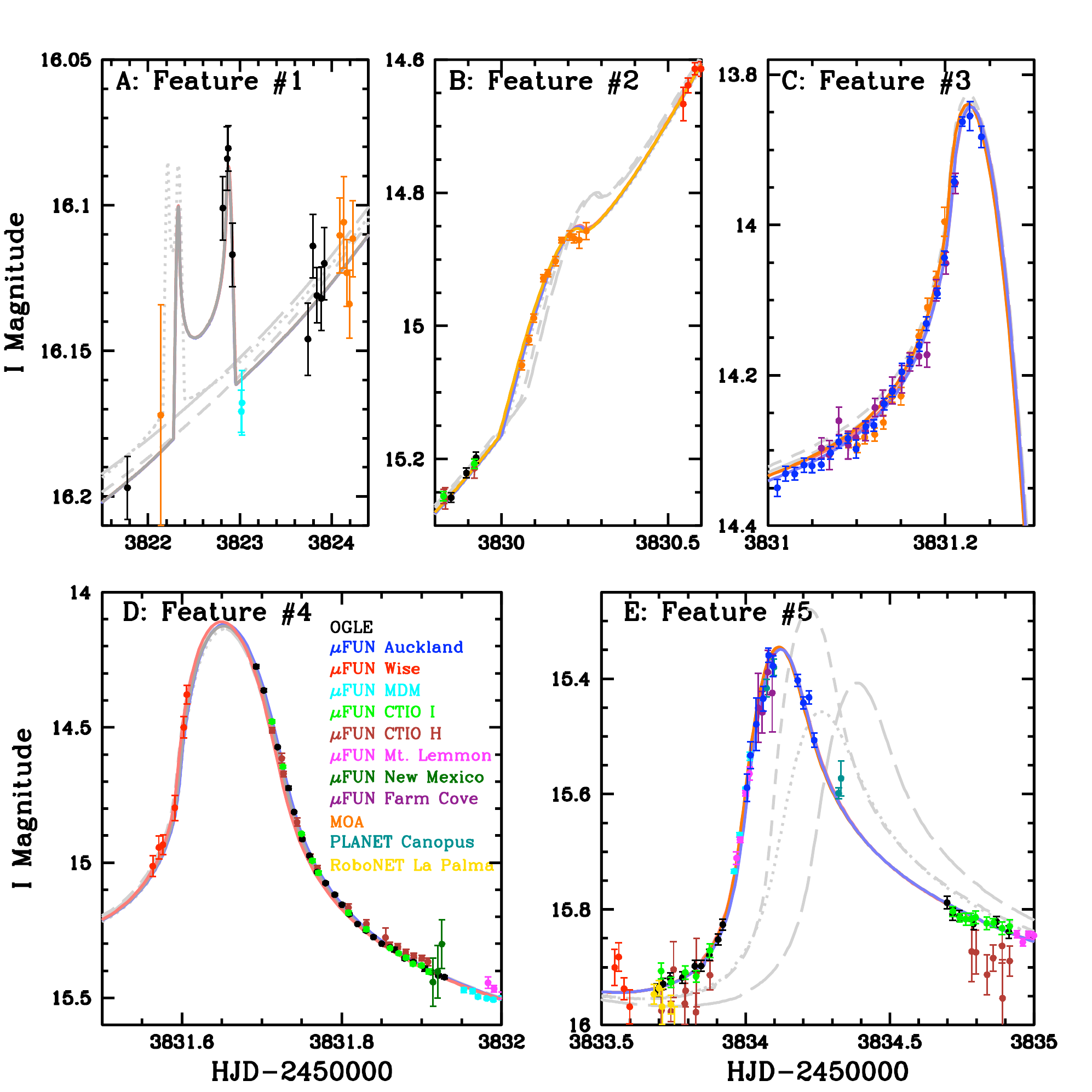}
\caption{Close-ups of the individual caustic crossing and cusp approach/crossing
features are plotted for the OGLE-2006-BLG-109 light curve. In this figure, it
is possible to make out the slight differences in the light curve 
due to the terrestrial
parallax effect, and so several different curves are plotted in different colors. The
light curve as seen from OGLE (in Las Campanas) is shown in gray,
the Wise (Israel) light curve is shown in red, the MOA (New Zealand
South Island) light curve is shown in orange, and the Auckland (New Zealand
North Island) is shown in blue. As in Figures~\ref{fig-lc1} and \ref{fig-lc2}, the
light gray curves plotted with dots, short-dashes, and long-dashes 
indicate the best-fit model without parallax, orbital motion, and both
parallax and orbital motion, respectively.
\label{fig-lc3}}
\end{figure}

\begin{figure}
\plotone{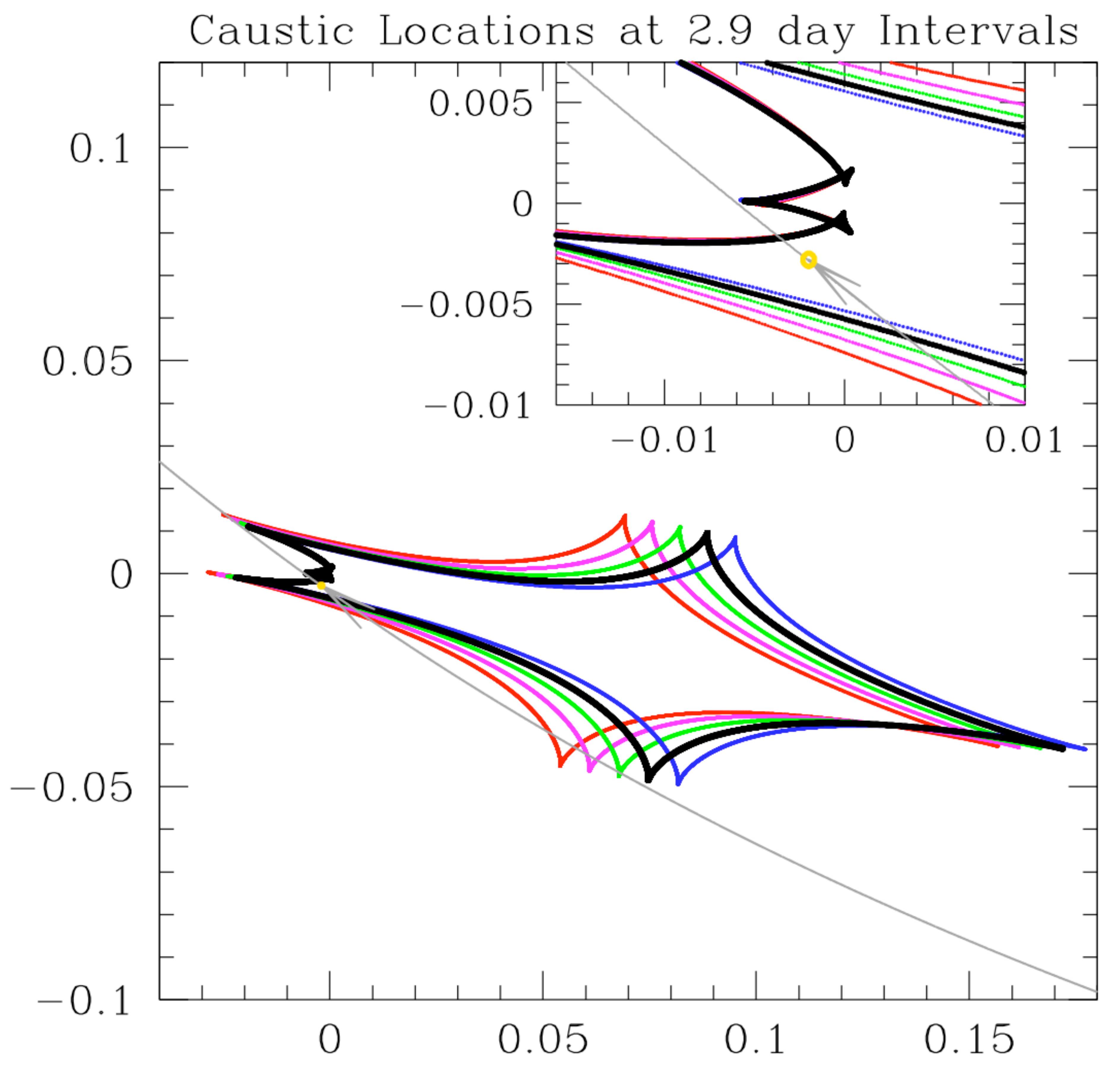}
\caption{Configuration of the central (resonant) caustic curves for 
OGLE-2006-BLG-109 is shown at five different times at intervals of
2.9 days ranging from $t = 3822.5$, which is the time of the first
caustic crossing to $t = 3834.1$, which is the time of the final cusp
approach. The red, magenta, green, black, and blue curves represent
$t = 3822.5$, 3825.4, 3828.3, 3831.2, and 3834.1, respectively.
The gray curve is the source trajectory, which is curved due to
microlensing parallax (\ie\ the motion of the Earth). The inset shows
a close-up of the central region of the central caustic, which includes the
triangular-shaped portion that is due to the Jupiter-mass planet,
OGLE-2006-BLG-109Lb. The gold circle indicates the angular size of
the source star.
\label{fig-caustic}}
\end{figure}

\begin{figure}
\plotone{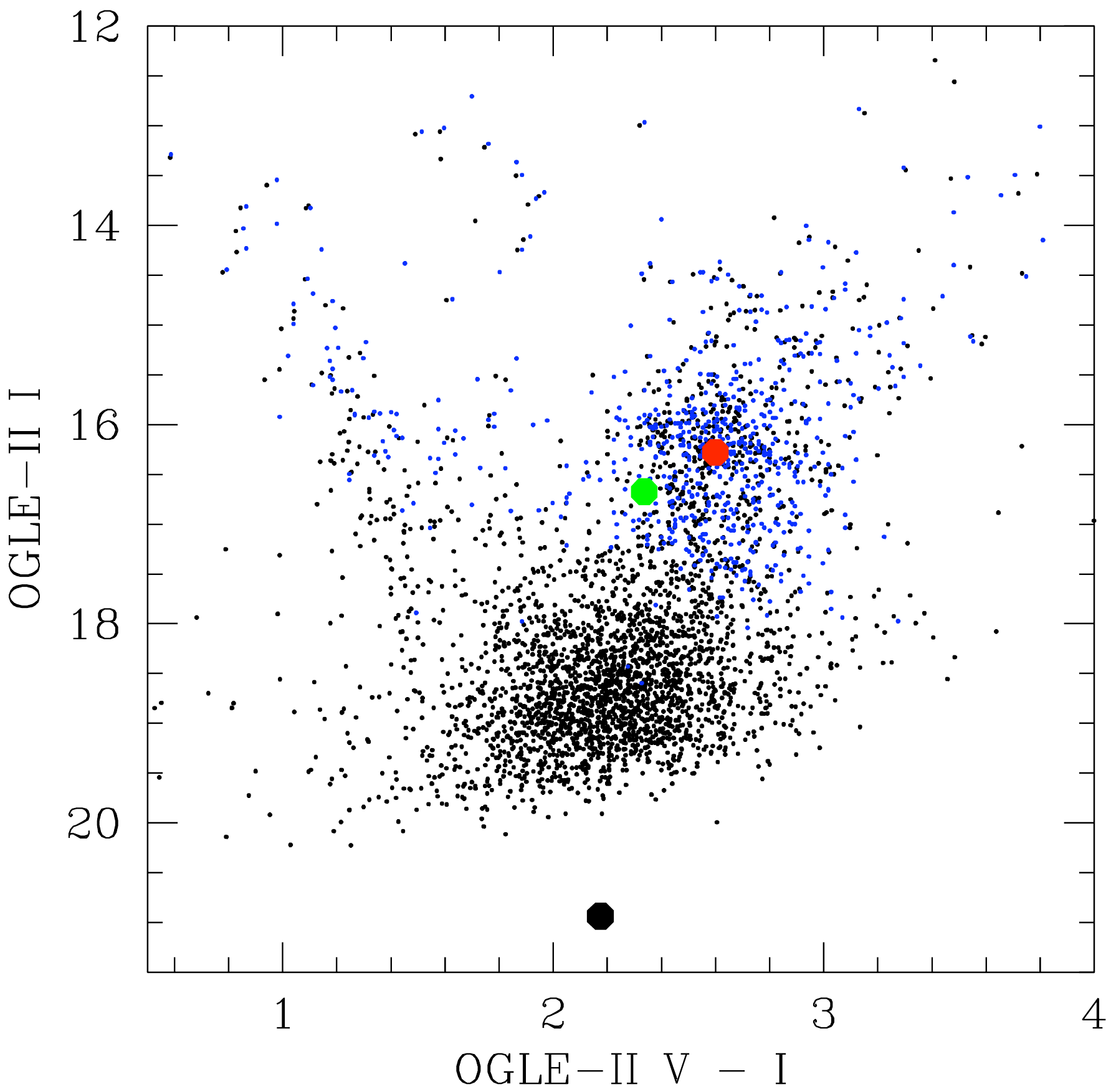}
\caption{CMD of the stars within
$60^{\prime\prime}$ of the OGLE-2006-BLG-109 source
star in $V$- and $I$-band magnitudes calibrated to the
OGLE-II system. The small black dots indicate the OGLE-II
photometry, and the blue dots indicate the CTIO photometry
that has been calibrated to the OGLE-II system and matched
with the CTIO $H$-band photometry (which does not go as deep
as $V$ and $I$).
The red dot indicates the inferred centroid
of the ``red clump"; the green dot indicates the location of
the bright star $0.35^{\prime\prime}$ from the source star,
and the large black dot indicates the location of the
source star.
\label{fig-cmd_VI}}
\end{figure}

\begin{figure}
\plotone{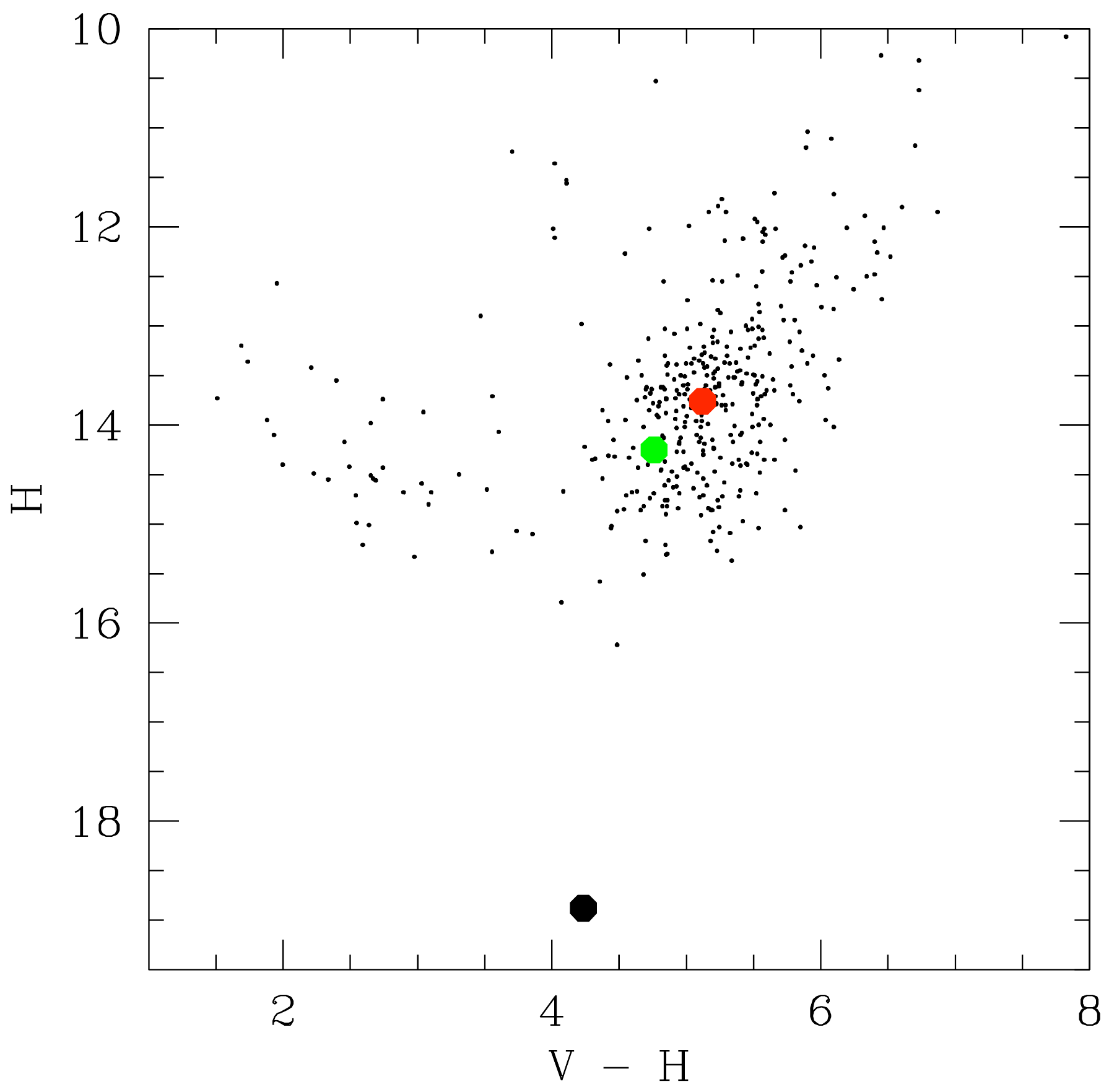}
\caption{$V-H$ CMD of the stars
within $60^{\prime\prime}$ of the OGLE-2006-BLG-109 source
star, based upon $V$- and $H$-band magnitudes from the 
Andicam instrument on the CTIO 1.3m telescope. 
The $V$ band has been calibrated to the
OGLE-II system, and the $H$ band has been calibrated
to 2MASS. As in 
Figure~\ref{fig-cmd_VI}, the red dot indicates the inferred centroid
of the ``red clump"; the green dot indicates the location of
the bright star $0.35^{\prime\prime}$ from the source star,
and the large black dot indicates the location of the
source star.
\label{fig-cmd_VH}}
\end{figure}

\begin{figure}
\plotone{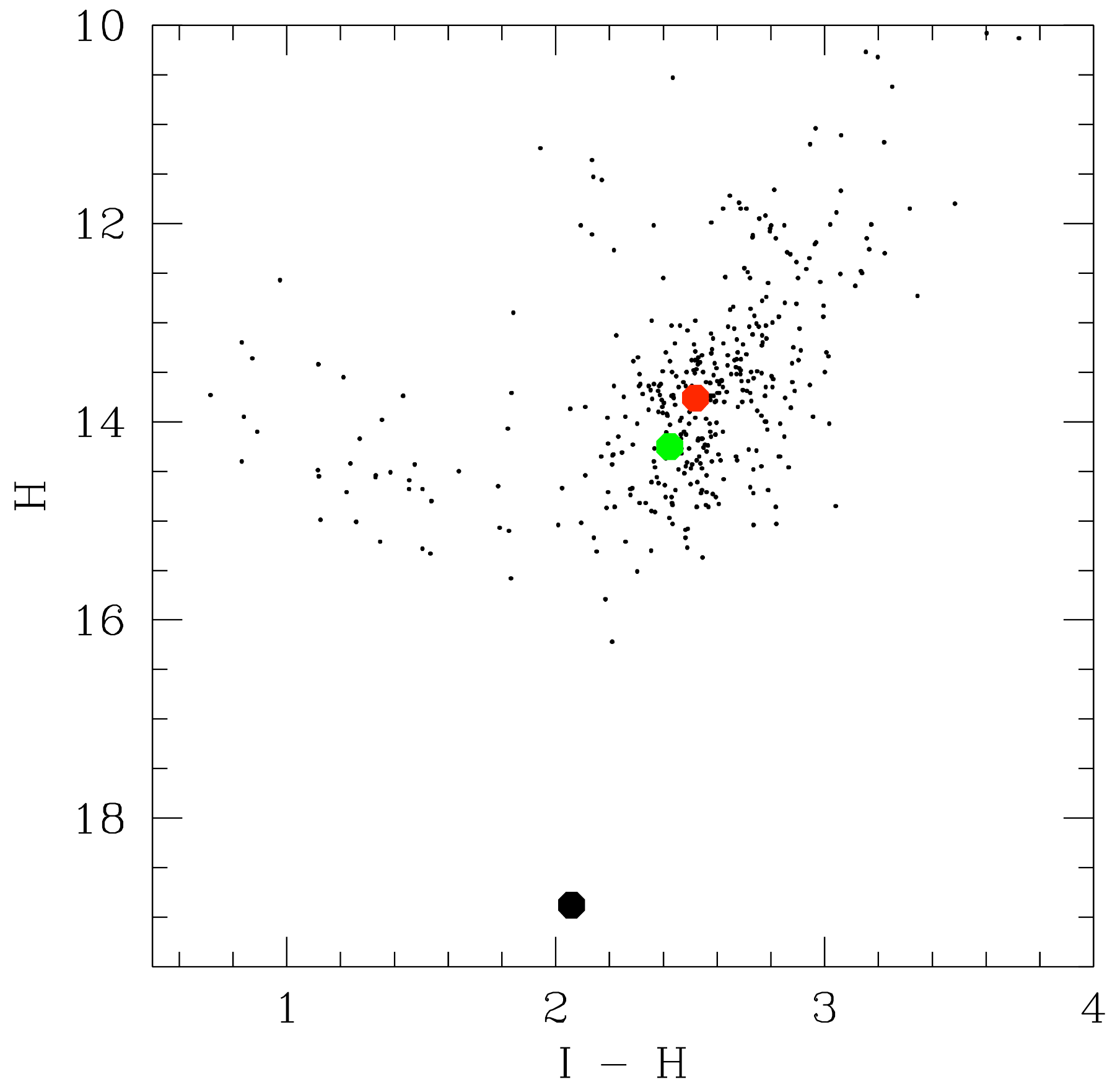}
\caption{$I-H$ CMD of the stars
within $60^{\prime\prime}$ of the OGLE-2006-BLG-109 source
star, based on the data described in Figures~\ref{fig-cmd_VI}
and \ref{fig-cmd_VH}.
\label{fig-cmd_IH}}
\end{figure}

\begin{figure}
\plotone{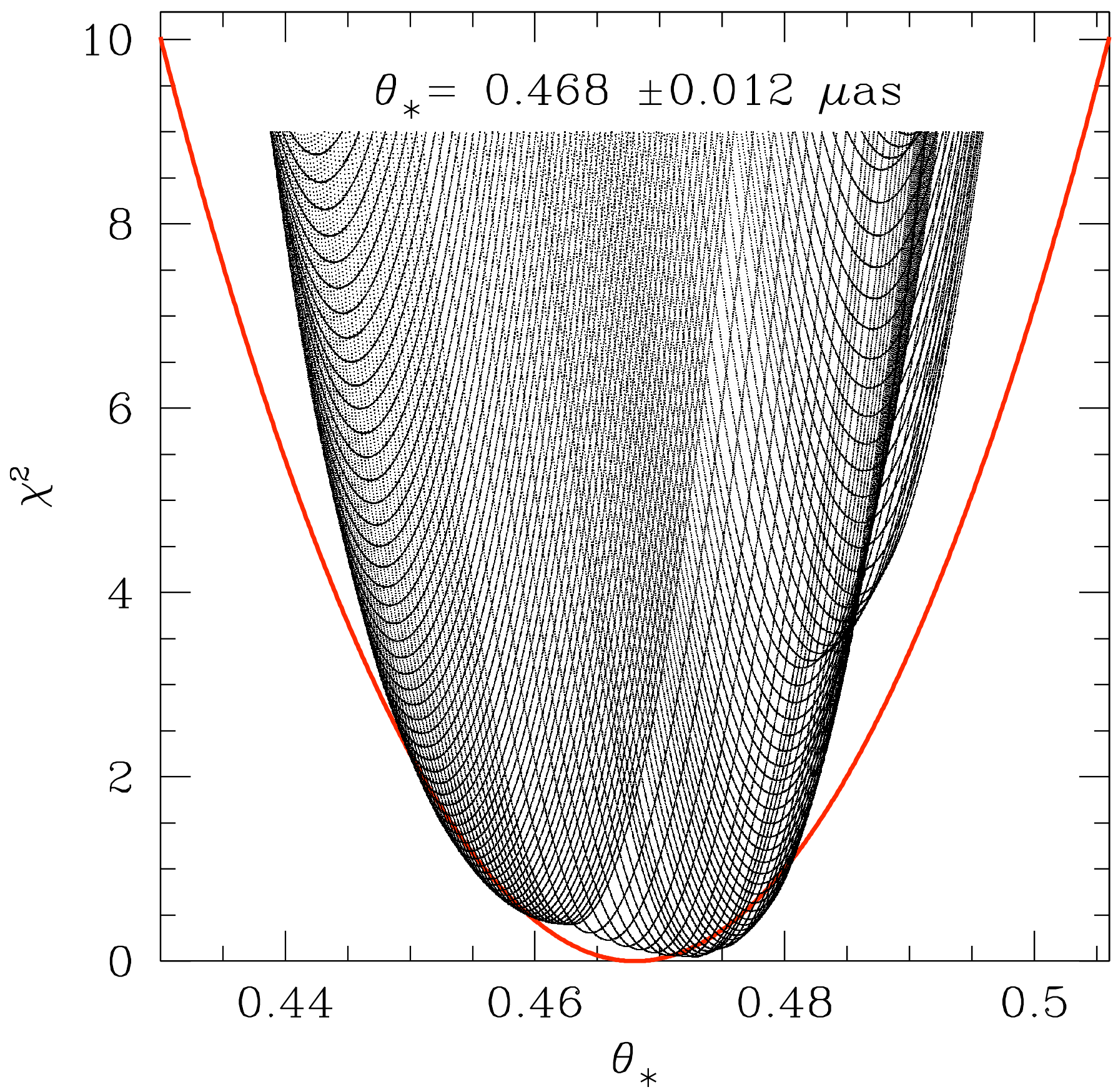}
\caption{$\chi^2$ vs.\ angular source radius, $\theta_\ast$, is plotted
for stars obeying the color-color relations of  \citet{kenyon_hartmann} subject
to extinction following the \citet{cardelli_ext} model using the 
$R_v$ distribution found for the red clump stars within $1^\prime$
of OGLE-2006-BLG-109S. The red curve is the $\chi^2$ parabola 
describing the result of our MCMC calculation: 
$\theta_\ast = 0.468 \pm 0.012\,\mu$as.
The $H$-band extinction is assumed to be within
5\% of the $A_{Hc} = 0.60$ value derived for this set of red clump stars.
\label{fig-radius}}
\end{figure}

\begin{figure}
\plotone{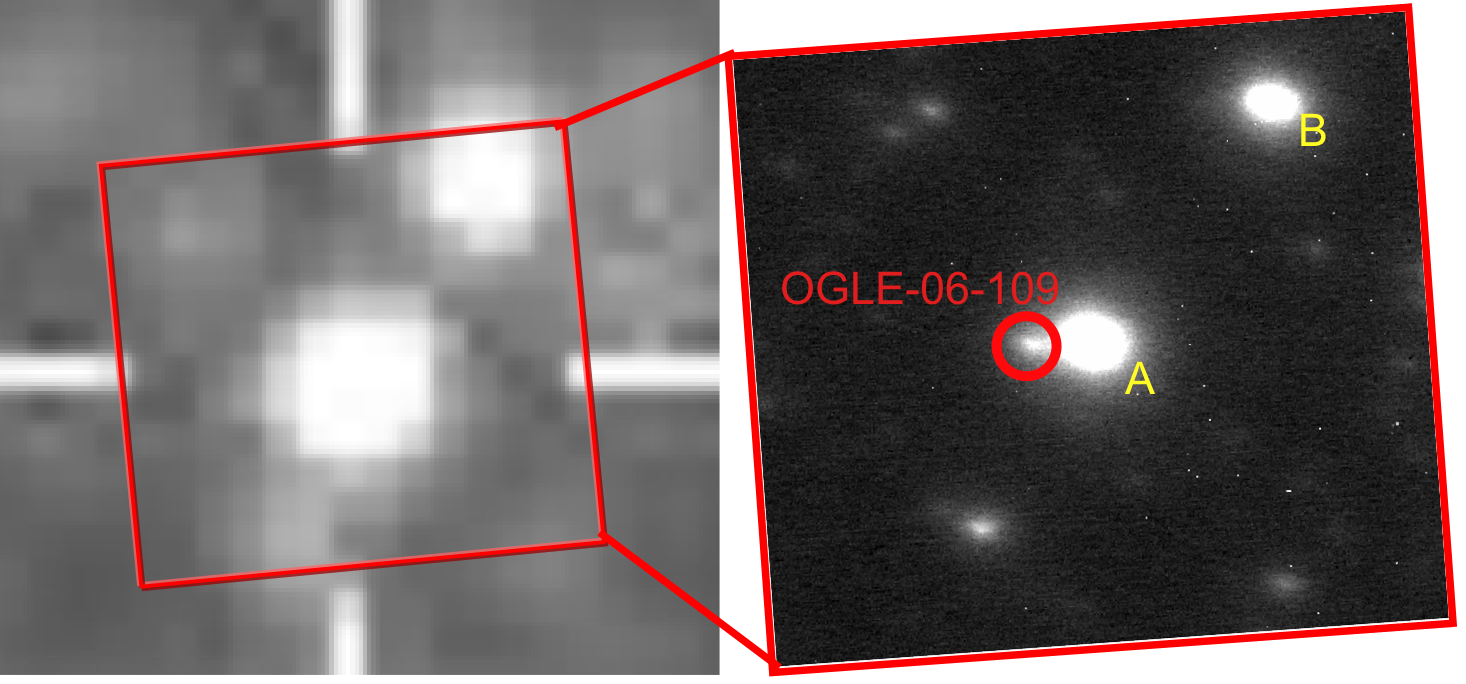}
\caption{Left panel shows a $6^{\prime\prime}\times 6^{\prime\prime}$
close-up of the OGLE $I$-band finding chart for OGLE-2006-BLG-109.
East is up and north is to the left. A first glance suggests that the source
is a red clump giant star, but a comparison of the astrometry
of the OGLE difference images and Keck images indicates that the source
is located $0.35^{\prime\prime}$ to the north of the red clump star seen in
the Keck images. 
The location of the OGLE-2006-BLG-109 source and lens stars is resolved in the
$3.8^{\prime\prime}\times 3.8^{\prime\prime}$ $H$-band Keck 
AO image on the right. 
\label{fig-keck}}
\end{figure}

\begin{figure}
\plotone{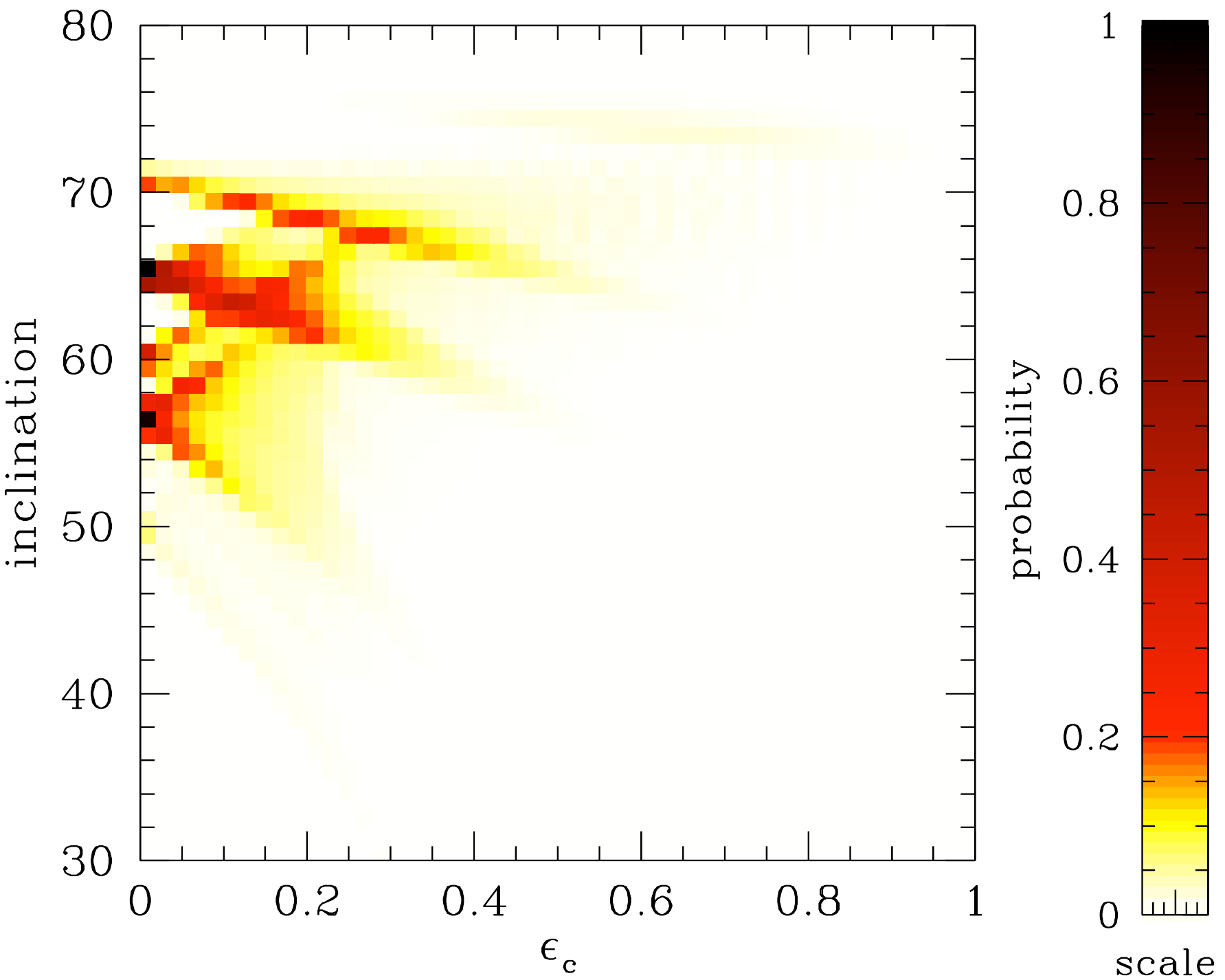}
\caption{Probability distribution for the orbit of the
Saturn-mass planet, OGLE-2006-BLG-Lc, is displayed in the 
eccentricity, $\epsilon_c$, vs.\ inclination angle, $i_c$,
plane. The sparse sampling of the $T_{\rm orb}$ parameter in the
MCMC calculations is partially responsible for the distinct 
high probability regions in parameter space.
\label{fig-eps_gam}}
\end{figure}

It seems likely that we will be able to extract even more information on 
exoplanetary systems from microlensing events discovered in the near future.
There have been two major telescope hardware improvements that have
occurred since the discovery of OGLE-2006-BLG-109 that should significantly
improve the light curve coverage for future events. Later in the 2006
observing season, the MOA Collaboration brought its 1.8m aperture
MOA-II telescope \citep{moa2_tel} online equipped with the $2.2\,{\rm deg}^2$
MOAcam-3 CCD camera \citep{sako_moacam3}. This allows an
observing cadence as high as one observation every 15 minutes for the 
central Galactic bulge fields. OGLE has just completed an upgrade to
its $1.4\,{\rm deg}^2$ OGLE-IV camera \citep{udalski_bohdan}
that will enable to increase its
sampling rate in a similar fashion. If OGLE-IV had been in operation when
OGLE-2006-BLG-109 was discovered, feature 1 in Figures~\ref{fig-lc1}
and \ref{fig-lc3} would have been recognized much sooner, which would
likely have resulted in many more observations of this feature. Similarly, 
the larger MOA-II telescope and higher observing cadence enabled by
MOA-cam-3 would have significantly improved the quantity and quality
of the MOA data. As a result, it is very likely that if a
similar event were detected
today, we would be able to tightly constrain the fifth orbital
motion parameter ($T_{\rm orb}$ in the parameterization we use here)
and to weakly constrain the final orbital parameter. This would yield
much tighter constraints on standard orbital parameters. So, for a fraction
of the future exoplanets discovered by microlensing, we can expect better
measurements of the orbital parameters than the ones we present
here for OGLE-2006-BLG-109.

\acknowledgments
We thank Ian Thompson and Andy McWilliam for
the reduction of the Magellan/MIKE spectrum of the lens star
and its bright neighbor.
D.P.B.\ was supported by grants 
AST-0708890 from the NSF and NNX07AL71G from NASA. 
The OGLE project is partially supported by the Polish MNiSW grant
N20303032/4275 to AU.
S.N.'s, B.M.'s and K.H.C.'s work was performed under the auspices of the U.S.
Department of Energy by Lawrence Livermore National Laboratory under
Contract DE-AC52-07NA27344.
T.S.\ was supported by grant JSPS18749004.
C.H.\ was supported by the National Research Fund og Korea 2009-0081561.
B-GP and C-UL were supported by Korea Astronomy and Space Science Institute.
Work by A.G. and S.D. was supported in part
by grant AST-0757888 from the NSF.
Work by B.S.G., A.G., and R.W.P was supported in part
by grant NNG04GL51G from NASA.

\clearpage


\clearpage

\begin{deluxetable}{cccc}
\tablecaption{Nonlinear Model Parameters
                         \label{tab-mparam} }
\tablewidth{0pt}
\tablehead{
\colhead{parameter}  & \colhead{units} &
\colhead{best fit value} & \colhead{MCMC range} 
}  

\startdata

$t_E$ & days & 127.300 & $128.1\pm 0.8$ \\
$t_0$ & ${\rm HJD}-2,450,000$ & 3831.0197 & $3831.0204\pm 0.011$ \\
$u_0$ & & 0.003479 & $0.00345\pm 0.00005$ \\
$d_{1\rm cm}$ & & 0.6272 & $0.632\pm 0.073$ \\
$d_{23}$ & & 1.04185 & $1.0418\pm 0.0001$ \\
$\theta_{1\rm cm}$ & radians & 2.52297 & $ 2.5232\pm 0.0007$ \\
$\phi_{23}$ & radians & -0.23560 & $-0.2350\pm 0.0007$ \\
$\epsilon_1$ & & $1.3562\times 10^{-3}$ & $(1.350\pm 0.013)\times 10^{-3}$ \\
$\epsilon_2$ & & $5.0516\times 10^{-4}$ & $(5.017\pm 0.030)\times 10^{-4}$ \\
$t_\ast$ & days & 0.03972 & $0.03949\pm 0.00016$ \\
$\dot{d}_{23x}$ & ${\rm days}^{-1}$ & 0.00169 & $0.00171\pm 0.00004$ \\
$\dot{d}_{23y}$ & ${\rm days}^{-1}$ & 0.00181 & $0.00179\pm 0.00014$ \\
$1/T_{\rm orb}$ & ${\rm yr}^{-1}$ & $2.04\times 10^{-4}$ & $2.3\pm 0.7\times 10^{-4}$\\
$\pi_E$ & & 0.3620 & $0.345\pm 0.014$ \\
$\phi_E$ & radians & 2.7296 & $2.728\pm 0.010$ \cr
fit $\chi^2$ & for 2557 dof & 2542.06 & \\

\enddata

\tablecomments{ Static parameters describe configuration
at ${\rm HJD}-2,450,000 = 3831.0$ (\ie, close to $t_0$). Mass 1
refers to the (Jupiter-mass) planet $b$; mass 2 is the (Saturn-mass) planet $c$
and mass 3 is the host star.}

\end{deluxetable}

\begin{deluxetable}{ccccccc}
\tablecaption{Physical Parameters
                         \label{tab-pparam} }
\tablewidth{0pt}
\tablehead{
 & & \multicolumn{5}{c} {parameter limits} \\
\colhead{parameter}  & \colhead{units} &
\colhead{$-2\sigma$} & \colhead{$-1\sigma$} &  \colhead{\bf median} &
\colhead{$+1\sigma$} & \colhead{$+2\sigma$} 
}  

\startdata

$D_L$ & kpc  & 1.30 & 1.39 & {\bf 1.51} & 1.62 & 1.74 \\
$M_A$ & $\msun $  & 0.43 & 0.47 & {\bf 0.51} & 056 & 0.60 \\
$m_b$ & $\mearth$   & 195 & 212 & {\bf 231} & 250 & 268 \\
$m_c$ & $\mearth$   & 73 & 79 & {\bf 86} & 93 & 99 \\
$M_{\rm H}$ & & 5.45 & 5.68 & {\bf 5.90} & 6.13 & 6.33 \\
$a_b$ & AU & 1.6 & 1.8 & {\bf 2.3} & 2.8 & 3.4 \\
$P_b$ & years & 2.8 & 3.4 & {\bf 4.9} & 6.5 & 7.3 \\
$a_c$ & AU & 2.9 & 3.5 & {\bf 4.5} & 6.6 & 13.5 \\
$P_c$ & years & 6.7 & 8.7 & {\bf 13.5} & 23.2 & 68 \\
$\epsilon_c$ & & 0.007 & 0.05 & {\bf 0.15} & 0.32 & 0.62 \\
$\alpha_c$ & deg. & -50 & -43 & {\bf -36} & -26 & -16 \\
$i_c$ & deg. & 49 & 56 & {\bf 64} & 68 & 73 \\
$K_b$ & m/sec & 14.6 & 16.3 & {\bf 17.4} & 18.7 & 19.9 \\
$K_c$ & m/sec & 2.8 & 3.9 & {\bf 4.5} & 5.0 & 5.3 \\
\enddata

\tablecomments{$M_H$ is the absolute $H$-band magnitude of the 
planetary host star. }

\end{deluxetable}

\begin{deluxetable}{ccccccc}
\tablecaption{Physical Parameters without Coplanar Assumption
                         \label{tab-pparam_noco} }
\tablewidth{0pt}
\tablehead{
 & & \multicolumn{5}{c} {parameter limits} \\
\colhead{parameter}  & \colhead{units} &
\colhead{$-2\sigma$} & \colhead{$-1\sigma$} &  \colhead{\bf median} &
\colhead{$+1\sigma$} & \colhead{$+2\sigma$} 
}  

\startdata

$a_c$ & AU & 2.5 & 3.3 & {\bf 4.8} & 8.4 & 15.6 \\
$P_c$ & years & 5.5 & 8.0 & {\bf 14.8} & 32 & 81 \\
$\epsilon_c$ & & 0.01 & 0.07 & {\bf 0.27} & 0.65 & 0.85 \\
$i_c$ & deg. & 37 & 51 & {\bf 62} & 72 & 75 \\
$K_c$ & m/sec & 2.6 & 3.6 & {\bf 4.2} & 5.0 & 5.3 \\
\enddata
\end{deluxetable}

\end{document}